\documentclass[fleqn,usenatbib]{mnras}

\usepackage{newtxtext,newtxmath}

\usepackage[T1]{fontenc}
\usepackage{ae,aecompl}

\usepackage{graphicx}	
\usepackage{amsmath}	
\usepackage{amssymb}	
\usepackage{times}
\usepackage{url}

\usepackage{color,ulem}

\newcommand{\removed}[1]{{\color{cyan}\it \sout{#1}}}

\title[Cavities in circumbinary discs]{On the cavity size in circumbinary discs}

\author[Hirsh, Price, Gonzalez, Ubeira-Gabellini \& Ragusa]
{Kieran Hirsh$^{1,2}$\thanks{E-mail: kieran.hirsh@univ-lyon1.fr},
Daniel J. Price$^{2}$, Jean-Fran\c{c}ois Gonzalez$^{1}$, M. Giulia Ubeira-Gabellini$^{3}$
\newauthor
and Enrico Ragusa$^{4}$
\\
$^{1}$Univ Lyon, Univ Claude Bernard Lyon 1, ENS de Lyon, CNRS, Centre de Recherche Astrophysique de Lyon UMR5574, F-69230, Saint-Genis-Laval, France\\
$^{2}$School of Physics and Astronomy, Monash University, VIC 3800, Australia\\
$^{3}$Dipartimento di Fisica, Universit{\`a} Degli Studi di Milano, Via Celoria, 16, Milano, I-20133, Italy.\\
$^{4}$School of Physics and Astronomy, University of Leicester, University Road, Leicester LE1 7RH, UK
}

\date{Accepted XXX. Received YYY; in original form ZZZ}

\pubyear{2020}
\defcitealias{AL94}{AL94}
\defcitealias{ML15}{ML15}
\begin{document}
\label{firstpage}
\pagerange{\pageref{firstpage}--\pageref{lastpage}}
\maketitle

\begin{abstract}
How does the cavity size in circumbinary discs depend on disc and binary properties? We investigate by simulating disc cavities carved by binary companions using smoothed particle hydrodynamics (SPH). We find that a cavity is quickly opened on the dynamical time, while the cavity size is set on the viscous time. In agreement with previous findings, we find long term cavity sizes of 2--5 times the binary semi-major axis, increasing with eccentricity and decreasing with disc aspect ratio. When considering binaries inclined with respect to the disc we find three regimes: i) discs that evolve towards a coplanar orbit have a large cavity, slightly smaller than that of an initially coplanar disc; ii) discs that evolve towards a polar orbit by breaking have a small cavity, equal in size to that of an initially polar disc; iii) discs that evolve towards a polar orbit via warping have an intermediate-sized cavity. We find typical gas depletions inside the cavity of $\gtrsim 2$ orders of magnitude in surface density.
\end{abstract}




\section{Introduction}

Recent spectacular resolved observations of cavities in the circumbinary discs HD142527 \citep{Casassus13,Avenhaus17}, and GG Tau \citep{Guilloteau99,Tang16,Yang17} and subsequent attempts to model them \citep{Cazzoletti17,Price18} have shown that binaries may be responsible for opening cavities that are large compared to the projected separation of the companion.

Indeed, even so-called `transitional discs' without detected companions, such as DZ Cha \citep{Briceno17,Canovas18}, DoAr 44 \citep{van-der-Marel16,Casassus18}, CQ Tau \citep{Tripathi17,Pinilla18,Margi19} and AB Aur \citep{Poblete20} contain features suggestive of a circumbinary disc, namely spiral arms and/or shadows around a central cavity.
Spiral arms occur in any disc with a companion \citep{OL02,Dong15a,Benisty17}, meanwhile shadows on the cavity edge or on the disc itself require some misalignment in the inner disc which can be caused by a companion on an inclined orbit \citep{Marino15,Min17}.
Cavities in transitional discs have been found to not be completely devoid of gas, but rather depleted in surface density by up to 5 orders of magnitude compared to the outer disc \citep{van-der-Marel15a,van-der-Marel16,van-der-Marel18}.
Finally, these discs show `horseshoes' or other asymmetries at the cavity edge \citep{Tuthill01,van-der-Marel13,Casassus15b} which may change on a timescale consistent with Keplerian motion of the outer disc \citep{Tuthill02}.
These asymmetries can be caused by the presence of a companion of planetary \citep{Ataiee13} or stellar \citep{Ragusa17} mass.
This model has been successfully applied to IRS 48 by \citet{Calcino19}.

The cavity opening process in circumbinary discs is a competition between the Lindblad resonances from the binary, which act to open a cavity, and the disc viscosity, which acts to close it (\citealt{AL94}, hereafter \citetalias{AL94}).
\citetalias{AL94} predicted a cavity size between 2--4 times the binary semi-major axis, becoming larger both with increasing binary eccentricity and decreasing disc viscosity.
Numerous computational studies have confirmed this basic picture \citep[\citetalias{AL94};][]{GK02,Thun17}, with some discrepancies over the exact cavity size.
\citet{Thun17}, for example, find a cavity size of up to seven times the binary semi-major axis. Their simulations, however, were evolved for 16,000 binary orbits, nearly 3 orders of magnitude longer than the original SPH simulation performed by \citetalias{AL94}.
However, both the latter studies were in 2D, limiting their applicability to coplanar discs.

\citet{ML15}, hereafter \citetalias{ML15}, generalised the analytical study by \citetalias{AL94} to discs inclined with respect to the binary orbital plane, and found that for prograde discs the cavity size tends to decrease with inclination.
To date however, few computational studies have considered the weakly inclined case, opting instead to consider the polar \citep{ML17,MarLub18,ML19} or retrograde \citep{NL15} cases.

In this paper we investigate what the observed cavity can tell us about unseen binary companions and the disc properties.
We perform a series of three-dimensional smoothed particle hydrodynamics (SPH) simulations to understand the effects of disc viscosity, binary eccentricity, binary mass ratio, and disc inclination on the cavity size.

\section{Cavity Opening in Circumbinary Discs}

For a binary of masses $M_1 \text{ and } M_2$, total mass $M_{\rm tot} = M_1 + M_2$, semi-major axis $a$, and eccentricity $e$, the orbital frequency is $\Omega_{\rm B} = (GM_{\rm tot}/a^3)^{1/2}$. The disturbing potential acting on the disc can be decomposed as a Fourier series, following the method outlined in \citetalias{ML15}, via
\begin{equation}
      \label{eq:disturb_pot}
      \Phi = \sum_{m,N}\Phi_{m,N}(r)\cos(m\phi - N\Omega_\text{B}t),
\end{equation}
where $(r,\phi)$ specify the radial and azimuthal positions of the disc particle at time $t$, $m$ is the azimuthal number in the disc, $N$ is the time harmonic number, and $\Phi_{m,N}(r)$ is the radial dependence of the potential component. Each $(m,N)$ component rotates with pattern frequency $\omega_{\rm P} = (N/m)\Omega_{\rm B}$ and excites density waves at the location of the Lindblad resonances (LRs), where the epicyclic frequency, $\kappa$, and forcing frequency are commensurate. Considering only the outer LRs, these are located where $\omega_{\rm P} - \Omega(r) = \kappa(r)/m$. If we assume a Keplerian disc then the LRs are located where $\Omega(r)/\Omega_{\rm B} = N/(m + 1)$, giving
\begin{equation}
      \label{eq:LR_loc}
      \frac{r_{\text{LR}}}{a} = \left(\frac{m + 1}{N}\right)^{\frac{2}{3}}.
\end{equation}

The torque at the $(m,N)$ LRs is given by \citep[e.g.][]{GT78}
\begin{equation}
      \label{eq:Lindblad}
      T^{\rm LR}_{m,N} = -m\pi^2\left[\Sigma\left(\frac{\text{d}D}{\text{d}\ln r}\right)^{-1}|\Psi_{m,N}|^2\right]_{r_\text{LR}},
\end{equation}
where $\Sigma$ is the disc surface density at the location of the resonance, $D = \kappa^2 - m^2(\Omega - \omega_{\rm P})^2$, and
\begin{equation}
      \label{eq:psi}
      \Psi_{m,N} = \frac{\text{d}\Phi_{m,N}}{\text{d}\ln r} + \frac{2\Omega}{\Omega-\omega_{\rm P}}\Phi_{m,N}.
\end{equation}
The viscous torque in the disc is given by \citep[e.g.][]{Pringle81}
\begin{equation}
      \label{eq:visc}
      T_\nu = 3\pi\alpha h^2\Sigma\Omega^2r^4,
\end{equation}
where $\alpha$ is the \citet{SS73} viscosity parameter, and $h = H/r$ is the disc aspect ratio. \citetalias{AL94} and \citetalias{ML15} assume that a gap will be opened at the $(m,N)$ LRs if
\begin{equation}
\label{eq:LR_vs_visc}
|T_\nu| \leq |T^{\rm LR}_{m,N}|.
\end{equation}
Since the binary torque exceeds the viscous one, the material is repelled from the cavity region on a dynamical timescale \citep[c.f:][]{Clarke01,AA07}, leaving behind a large cavity out to the farthest LR satisfying Equation~\ref{eq:LR_vs_visc}.

\section{Methods}

\subsection{Initial Conditions}

Using the SPH code \textsc{Phantom} \citep{PHANTOM} we model a gas disc consisting of one million particles initially placed in a circumbinary disc extending from $1.4$ to $14.5$ times the binary semi-major axis, with the binary modelled as a pair of sink particles following the prescription of \citet{Bate95}.
We simulate binaries with mass ratios of $q = 0.01, 0.1, 0.3\text{ and }0.5$ with $q = M_2/(M_1 + M_2)$, where $M_1$ and $M_2$ are the mass of the primary and secondary, respectively.
We use a disc mass of $M_{\rm{disc}} = 0.0001 M_1$, in order to reduce the effects of the disc gravity on the binary orbit.
This low mass leads to a negligible disc self-gravity, so we do not include it in our simulations.
We assume a surface density profile $\Sigma \propto R^{-p}$, with $p = 1.0$.
We prescribe a locally isothermal equation of state, that is $P = c_s^2(R)\rho$, with sound speed varying as $c_s \propto R^{-w}$, with $w = 0.25$.
This leads to a temperature profile $T \propto R^{-2w}$ and a disc aspect ratio varying as $H/R \propto R^{1/2 - w}$.
This allows us to set the sound speed, temperature and aspect ratio by specifying the aspect ratio at the disc inner edge. We simulate discs with $(H/R)_{\rm{in}} = 0.01, 0.02, 0.04, 0.05, 0.06, 0.08, 0.10\text{ and }0.12$.
The setup for our fiducial simulation, as well as the full parameter space investigated, is outlined in Table~\ref{table:standard_sim}.

\begin{table}
      \centering
      \caption{Simulation parameters. We vary the binary mass ratio, disc inclination, and disc scale height. Varying the scale height corresponds to the value of artificial viscosity given beneath it.}
      \begin{tabular}{@{}lcc@{}} 
            \hline
            Parameter & Fiducial value & Other explored values \\
            \hline
            $q$ & 0.1 & 0.01, 0.3, 0.5 \\
            $M_{\rm disc}/M_1$ & $0.0001$ \\
            $R_{\rm in}/a$ & 1.4 \\
            $R_{\rm out}/a$ & 14.5 \\
            $p$ & 1.0 \\
            $w$ & 0.25 \\
            $\alpha$ & 0.005 \\
            $\rm{inclination}$ & $0^{\circ}$ & $22.5^{\circ}$, $45^{\circ}$, $90^{\circ}$ \\
            \hline
            \multicolumn{3}{@{}l}{Viscosity-dependent parameters} \\
            \hline
            $(H/R)_{\rm{in}}$ & 0.05 & 0.01, 0.02, 0.04, 0.06, 0.08, 0.10, 0.12 \\
            $\alpha_{\rm AV}$ & 0.20 & 0.07, 0.11, 0.17, 0.22, 0.27, 0.31, 0.35 \\
            \hline
      \end{tabular}
      \label{table:standard_sim}
\end{table}

\begin{figure*}
	\includegraphics[width=\textwidth]{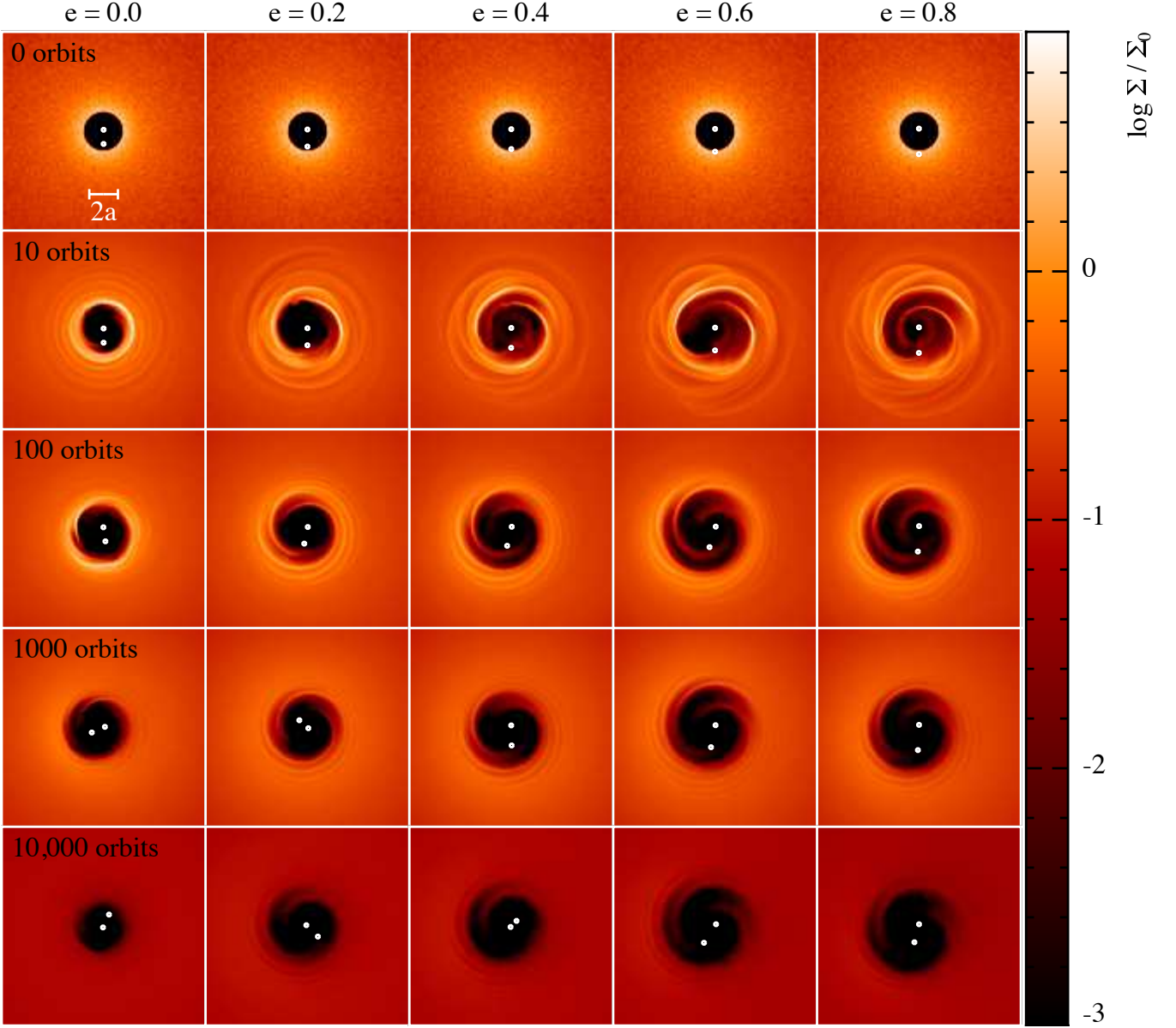}
    \caption{Surface density rendered face-on views of the evolution of coplanar discs with $(H/R)_{\rm in} = 0.05$ surrounding a binary with $q = 0.1$. Eccentricity increases from left to right and time increases from top to bottom. 
    }
    \label{fig:q01_i000_HR05_cav_vs_eccen_rendered}
\end{figure*}

\subsection{Disc Viscosity}
\label{sec:disc_visc}

We prescribe an $\alpha$ disc, i.e. the disc viscosity is $\nu = \alpha c_s H$ \citep{SS73}, and model the disc viscosity as in \citet{LP10} using the SPH artificial viscosity parameter, \(\alpha_\text{AV}\) which can be related to the Shakura-Sunyaev $\alpha$ using
\begin{equation}
      \label{eq:disc_visc}
      \alpha \simeq \frac{\alpha_\text{AV}}{10}\frac{\langle h\rangle}{H},
\end{equation}
where $H$ is the scale height of the disc and $\langle h\rangle$ is the azimuthally averaged smoothing length.
Since $H \equiv c_s/\Omega$ our expression for the viscosity can be rewritten as $\nu = (\alpha_\text{AV}/10)\langle h\rangle H \Omega$.
By setting $\alpha_\text{AV}$ such that the average $\alpha = 0.005$ we can then vary the viscosity by varying the scale height of the disc.

The corresponding viscous time $t_\text{visc} = R^2/\nu$, at $R = R_{\rm{in}}$, is given in terms of the orbital time ($2\pi/\Omega$) according to
\begin{equation}
      \label{eq:visc_time}
      t_\text{visc} \approx 12,800 \text{ orbits} \left(\frac{\alpha}{0.005}\right)^{-1} \left(\frac{H/R}{0.05}\right)^{-2}.
\end{equation}
For the discs we investigate this gives a $t_\text{visc}$ that varies from roughly $2,200$ orbits for $(H/R)_{\rm{in}} = 0.12$, to roughly $320,000$ orbits for $(H/R)_{\rm{in}} = 0.01$.
Physically it is more sensible to consider $t_\text{visc}$ at the cavity edge ($R_{\rm cav}$), but since this varies throughout and between simulations we consider $t_\text{visc}$ at $R_{\rm{in}}$ and note a discrepancy of a factor of $R_{\rm{cav}}/R_{\rm{in}}$.

\subsection{Cavity Size}

We azimuthally average the surface density and define the half-maximum radius to be the radius at which the surface density first reaches half its maximum, with a similar definition for the quarter-maximum density.
Following the prescription in \citetalias{AL94}, we then take the cavity size to be the radius at half-maximum, with a symmetric error taken as the difference between the radii at half-maximum and quarter-maximum.

\begin{figure}
   \centering
      \includegraphics[width=0.5\textwidth]{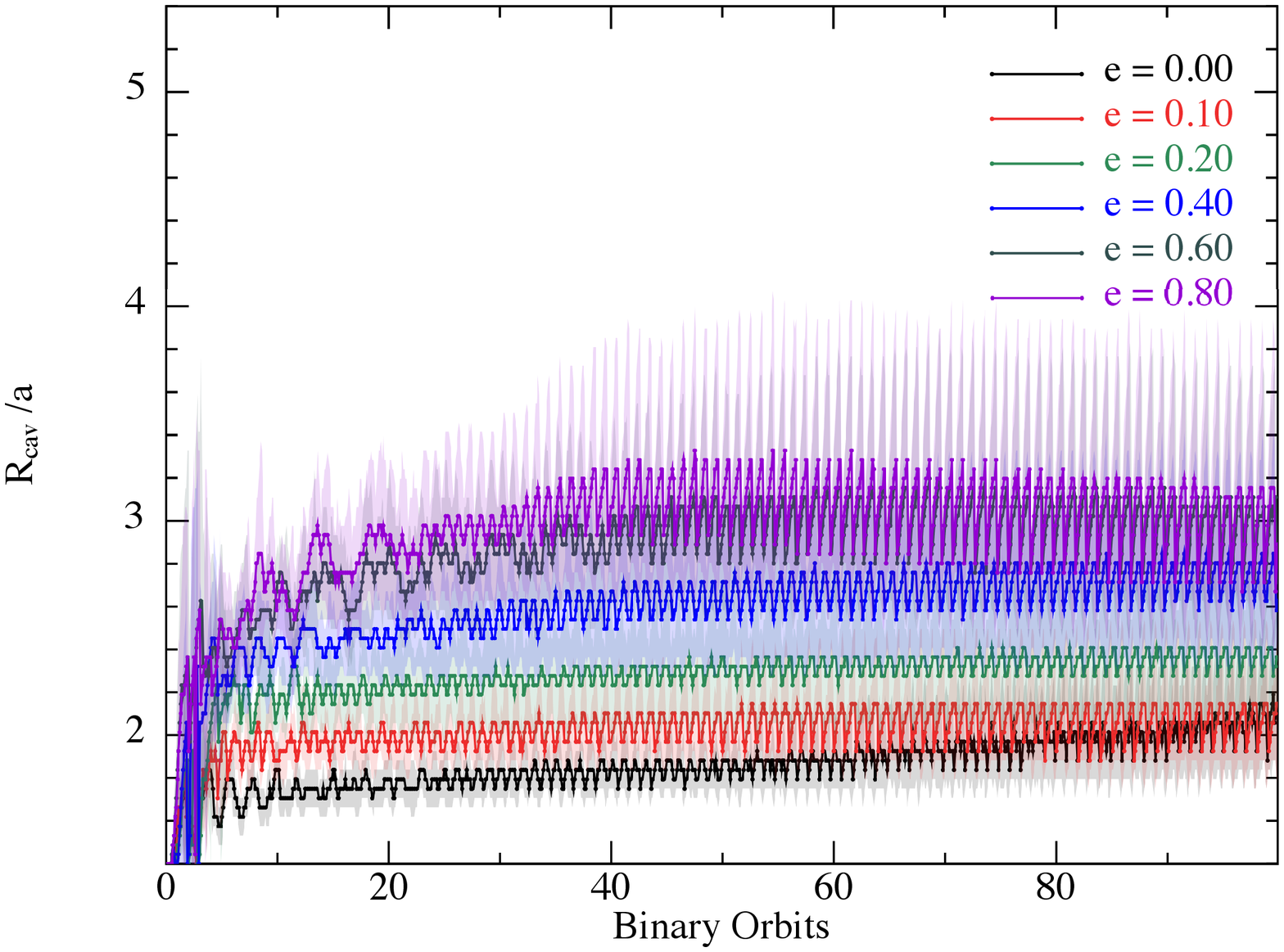}
      \includegraphics[width=0.5\textwidth]{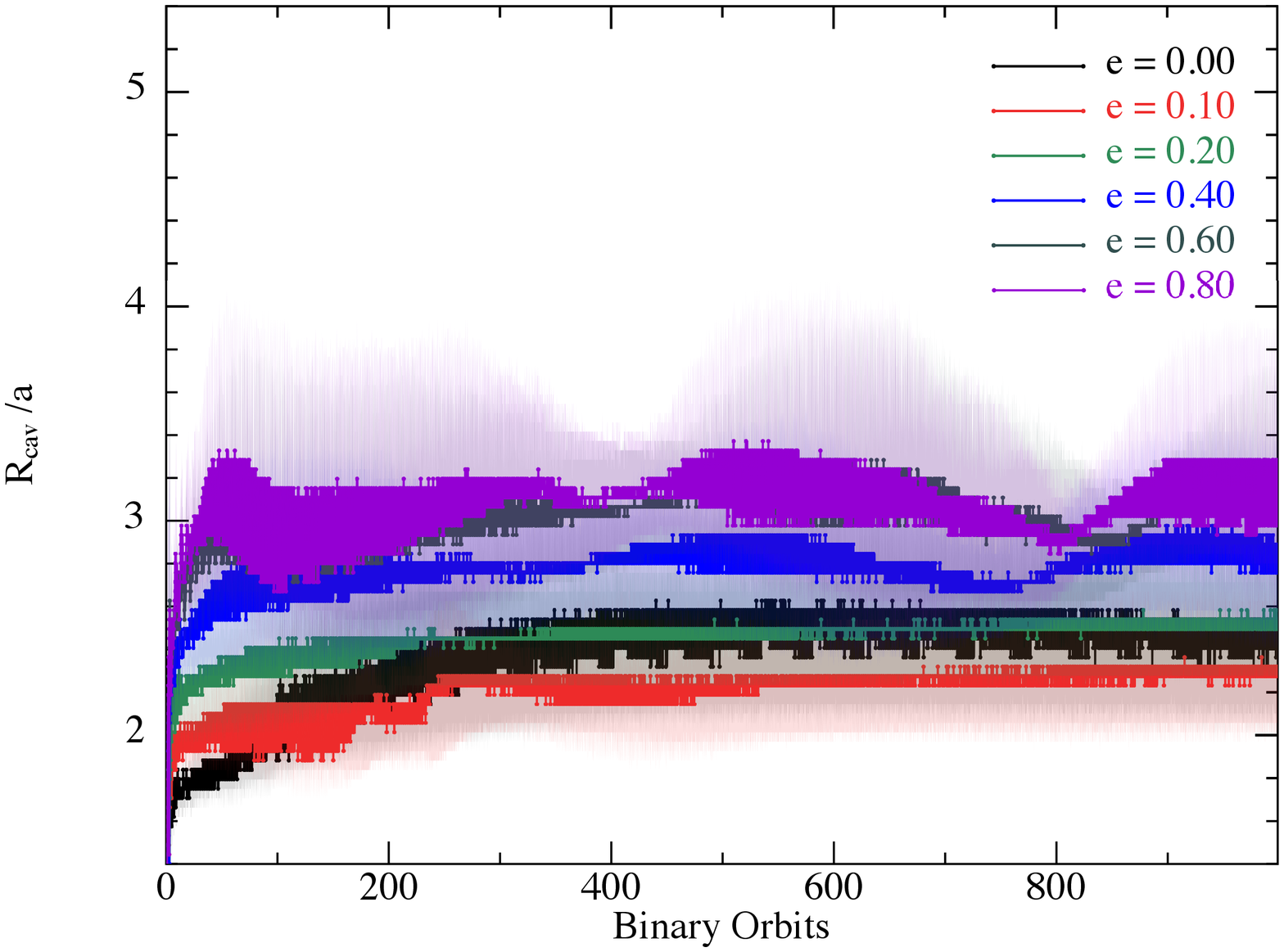}
      \includegraphics[width=0.5\textwidth]{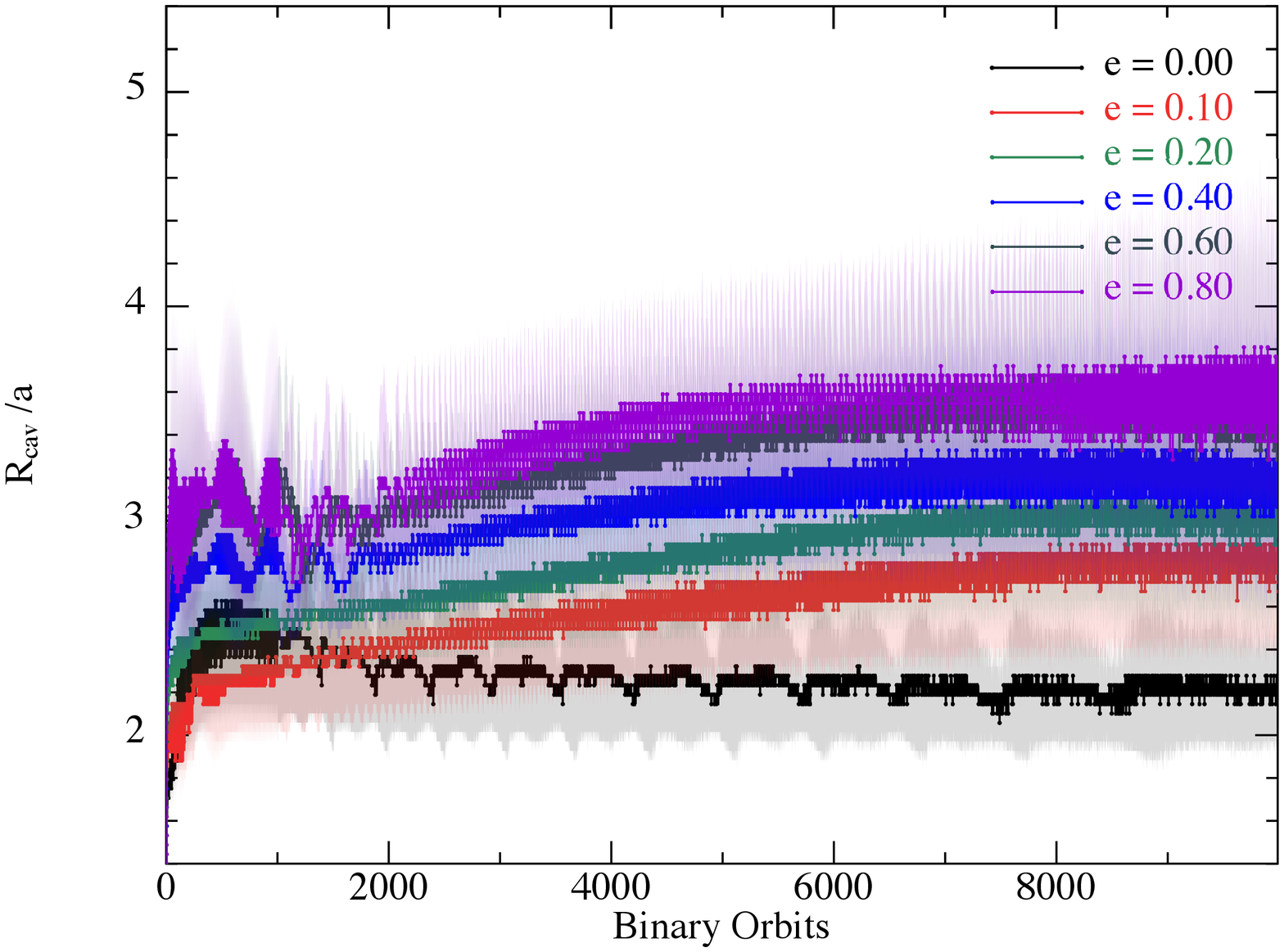}
   \caption{Time evolution of the cavity size for a coplanar disc with $(H/R)_{\rm{in}} = 0.05$ surrounding a binary with $q = 0.1$ over 100 binary orbits (top panel), 1000 binary orbits (middle panel) and 10,000 binary orbits (bottom panel), for different initial binary orbital eccentricities (see legend). The shaded region represents the error bars in measurement of $R_{\rm cav}$, as is the case for all subsequent plots in this paper.
   }
   \label{fig:q01_i000_HR05_cav_vs_time}
\end{figure}

\section{Results}

\subsection{Time Evolution}
\label{sec:time_evol}

Figure~\ref{fig:q01_i000_HR05_cav_vs_eccen_rendered} shows surface density rendered face-on views of the cavity opening process for a coplanar disc with $(H/R)_{\rm in}=0.05$ and $q = 0.1$ and eccentricities ranging from $e=0$ to $e=0.8$.
The cavity size increases with time (top to bottom) until reaching an equilibrium after several thousand orbits.
After 10,000 binary orbits the background surface density is smaller due to viscous disc spreading.

Figure~\ref{fig:q01_i000_HR05_cav_vs_time} quantifies the cavity size as a function of time and initial binary eccentricity.
The top panel shows the evolution on tens of dynamical timescales (a dynamical timescale being $\lesssim10$ binary orbits at $R_{\rm{cav}}$).
The cavity is opened on this timescale and the size appears to stabilise between $2-3$ times the semi-major axis depending on the eccentricity of the binary.
Evolving the system on the viscous timescale ($\sim10,000$ binary orbits) shows the cavity continue to grow to $2.5-3.5$ times $a$ for eccentric binaries (bottom panel).
The circular case is unique in that it reaches a maximum cavity size of the order of hundreds of binary orbits, while eccentric binaries continue to grow their cavities for thousands of binary orbits.

\subsection{Binary Orbital Eccentricity}
\label{sec:binary_eccen}

\begin{figure}
	\includegraphics[width=0.5\textwidth]{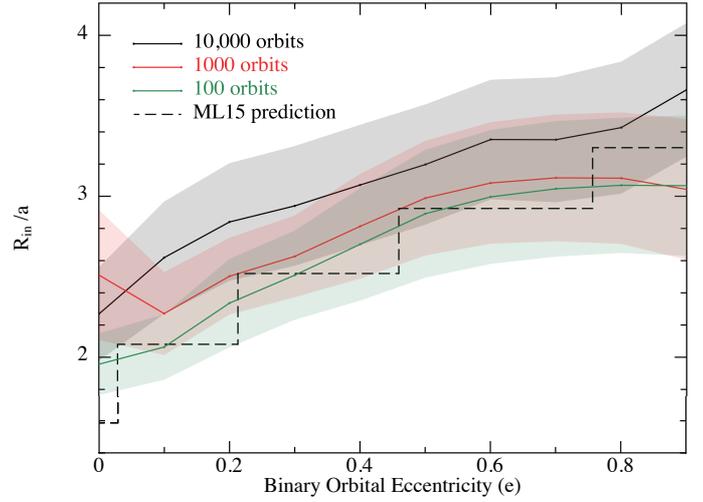}
    \caption{Cavity size as a function of initial binary orbital eccentricity for a coplanar disc with $(H/R)_{\rm{in}} = 0.05$ surrounding a binary with $q = 0.1$. Snapshots are taken after 100 (green line), 1000 (red line) and 10,000 (black line) binary orbits. Dashed line shows prediction from \citet{ML15}.
    }
    \label{fig:q01_i000_HR05_cav_vs_eccen}
\end{figure}

Figure~\ref{fig:q01_i000_HR05_cav_vs_eccen_rendered} shows the effect of binary eccentricity on the cavity size.
Cavity size increases with increasing eccentricity.
This is shown quantitatively in Fig.~\ref{fig:q01_i000_HR05_cav_vs_eccen}.
At early (100 binary orbits; green line) and late (10,000 binary orbits; black line) stages the cavity size increases with binary orbital eccentricity, consistent with both \citetalias{AL94} and \citetalias{ML15}.
After 1000 binary orbits (red line), however, we see a turnover in the cavity size due to the circular binaries reaching a maximum cavity size before eccentric ones.
This turnover is only temporary though, and disappears once the eccentric binaries reach a maximum cavity size.
\citet{Thun17} also find a turnover in the cavity size, however theirs persists up to 16,000 binary orbits, and the minimum is seen at $e \approx 0.18$ while ours is at $e \approx 0.1$.

The exact values for the cavity size also show some discrepancies.
\citet{Thun17} found cavity sizes between 4 and 7 times the binary semi-major axis, nearly double the values found by our work, as well as that of \citetalias{AL94} and \citetalias{ML15} (dashed lines in Fig.~\ref{fig:q01_i000_HR05_cav_vs_eccen}).
We discuss this difference in Section~\ref{sec:discussion}.

\subsection{Disc Scale Height}
\label{Sec:scale_height}

\begin{figure*}
	\includegraphics[width=\textwidth]{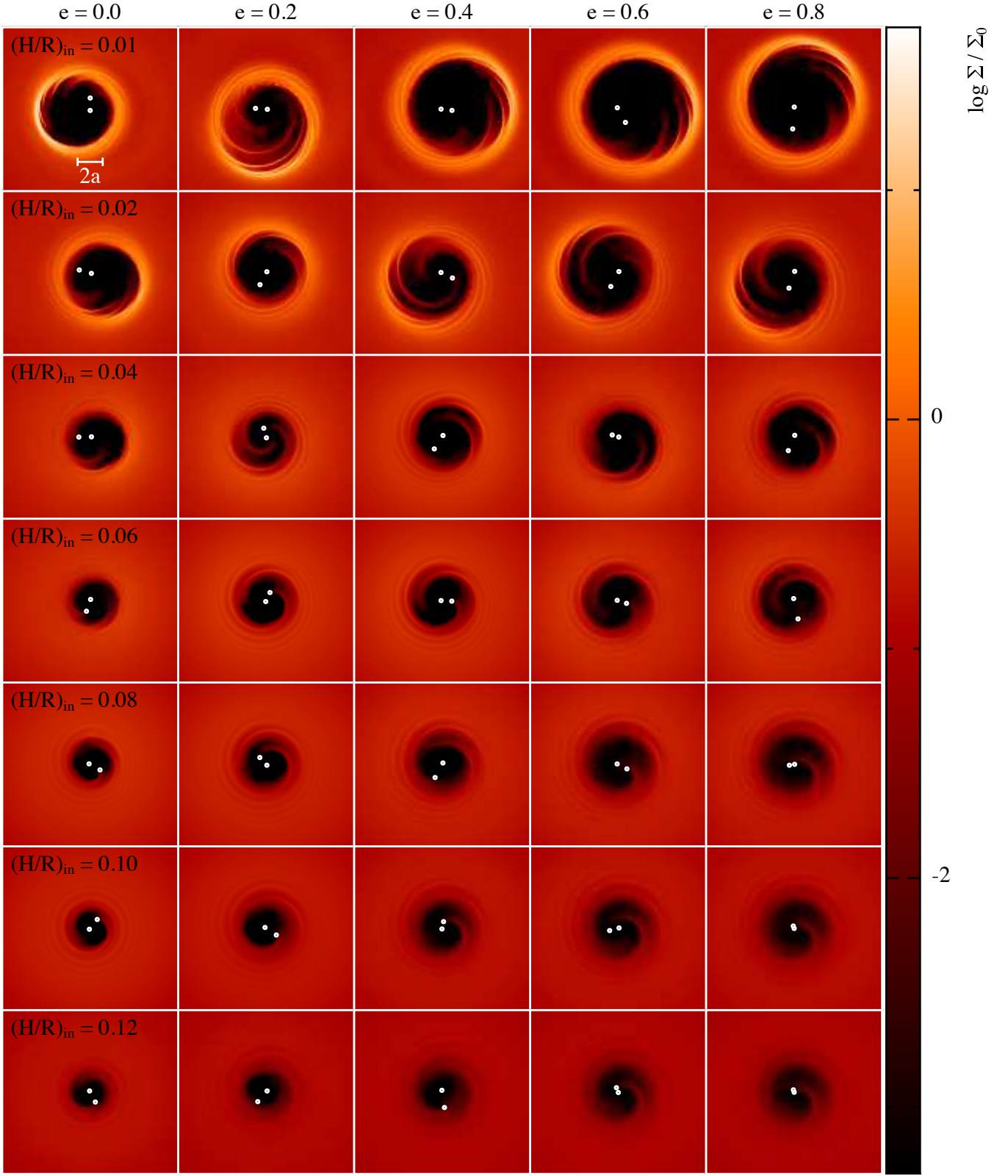}
    \caption{Surface density rendered face-on views of coplanar circumbinary discs surrounding a binary with $q = 0.1$ after 1000 binary orbits. Disc aspect ratio increases top to bottom, initial binary orbital eccentricity increases left to right. As seen in Figs.~\ref{fig:q01_i000_HR05_cav_vs_eccen}~and~\ref{fig:q01_i000_cav_vs_HR_multiplot} cavity size increases with binary orbital eccentricity and decreases with disc aspect ratio.
    }
    \label{fig:q01_HR_rendered}
\end{figure*}

Figure~\ref{fig:q01_HR_rendered} shows the surface density rendered face-on views of discs evolved for 1000 binary orbits with various eccentricities (increasing left to right) and disc scale heights (increasing top to bottom).
We see the cavity size increase with binary eccentricity, as described in Section~\ref{sec:binary_eccen}, and decrease with increasing scale height.
We also see the most eccentric cavities around the discs with smallest scale height.

\begin{figure}
	\includegraphics[width=0.5\textwidth]{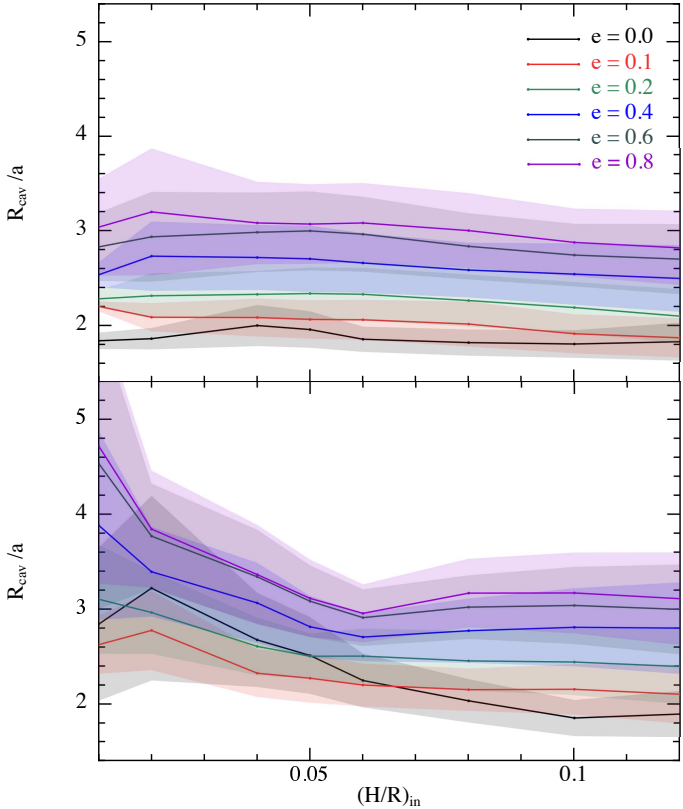}
    \caption{Cavity size as a function of disc aspect ratio for a coplanar disc surrounding a binary with mass ratio $q = 0.1$ after 100 binary orbits (top panel) and 1000 binary orbits (bottom panel).
    }
    \label{fig:q01_i000_cav_vs_HR_multiplot}
\end{figure}

Care must be taken, however, to evolve the discs for a significant fraction of the viscous time.
The top panel of Fig.~\ref{fig:q01_i000_cav_vs_HR_multiplot} shows the cavity size as a function of disc aspect ratio after only 100 binary orbits.
From Equation~\ref{eq:visc_time}, this corresponds to $\sim 3 \times 10^{-4} t_\text{visc}$ for $(H/R)_{\rm{in}} = 0.01$ and $\sim 4.5 \times 10^{-2} t_\text{visc}$ for $(H/R)_{\rm{in}} = 0.12$.
At this early stage there is no dependence of cavity size on disc aspect ratio.
The bottom panel of Fig.~\ref{fig:q01_i000_cav_vs_HR_multiplot} is the same as the top panel, but after $1000$ binary orbits.

Although 1000 binary orbits does not fully resolve the viscous time, it is already possible to see trends appearing.
When $(H/R)_{\rm{in}} \lesssim 0.06$ the cavity size decreases for increasing scale height, then remains largely unchanged above this value.
Furthermore, while the most viscous discs with $(H/R)_{\rm{in}} \gtrsim 0.06$ continue to evolve after 100 orbits the change in cavity size is minor, remaining within error bars.
This suggests that taking the cavity size after 1000 orbits ($\gtrsim 0.1 t_{\rm visc}$ for these highly viscous cases) provides a reasonable estimation of the long-term cavity size.

While longer simulations would allow us to fully resolve the viscous time, these simulations become prohibitively expensive at low viscosity, requiring more than $10^5$ binary orbits for $(H/R)_{\rm{in}} = 0.01$.
It is also important to note that such long simulations would reach, or even exceed, the expected lifetime of protoplanetary discs, reducing their applicability to planet-forming discs at these late times.

\begin{figure*}
	\includegraphics[width=\textwidth]{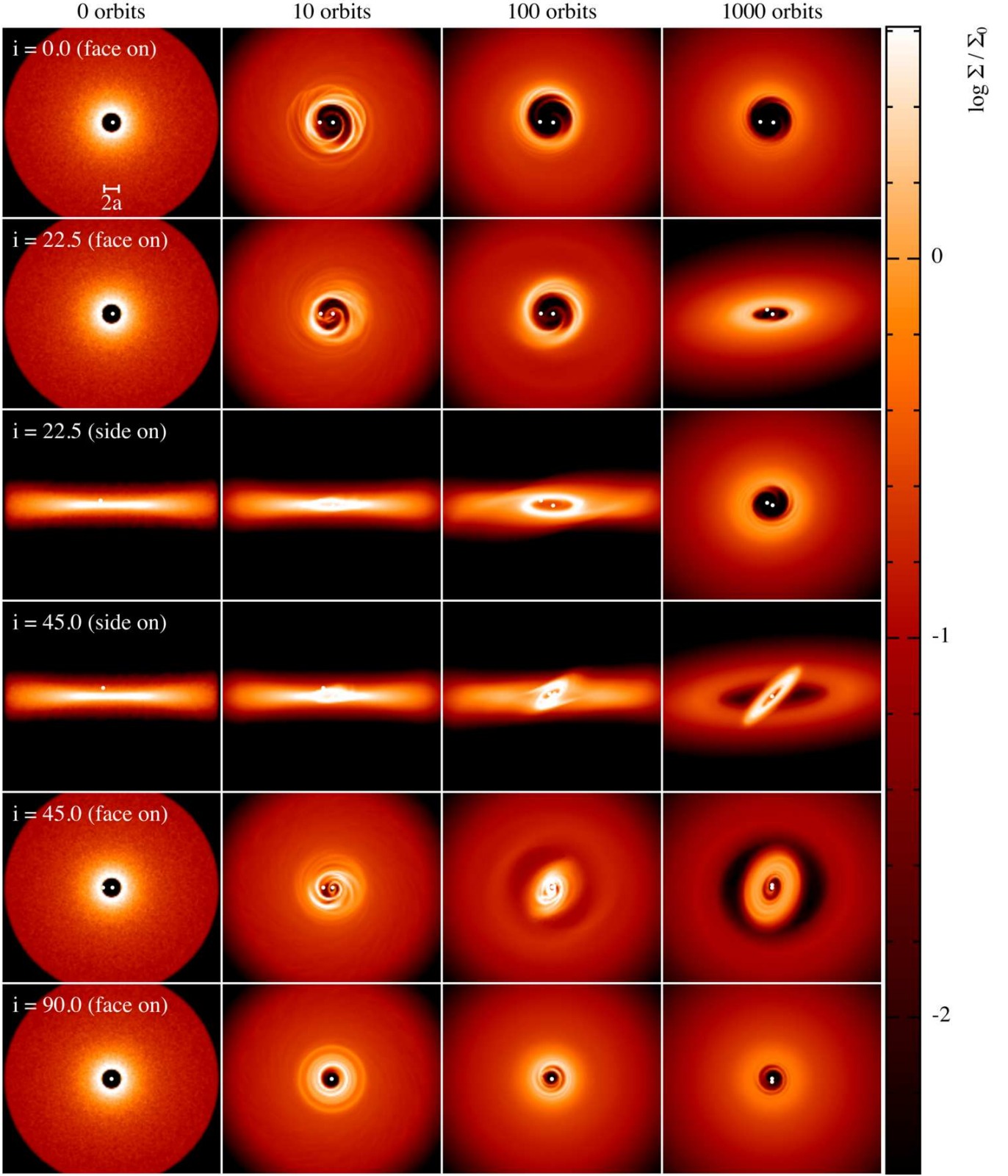}
	\caption{Surface density rendered views of the evolution of inclined circumbinary discs with $(H/R)_{\rm{in}} = 0.05$ surrounding a binary with $q = 0.1$ and $e = 0.8$. Time increases from left to right and inclination increases from top to bottom. Coplanar and polar discs are shown only in face-on views, while $i = 22.5^\circ$ and $i = 45^\circ$ are shown in both face-on and side-on views.
    }
    \label{fig:q01_incl_rendered_both}
\end{figure*}

\begin{figure}
	\includegraphics[width=0.45\textwidth]{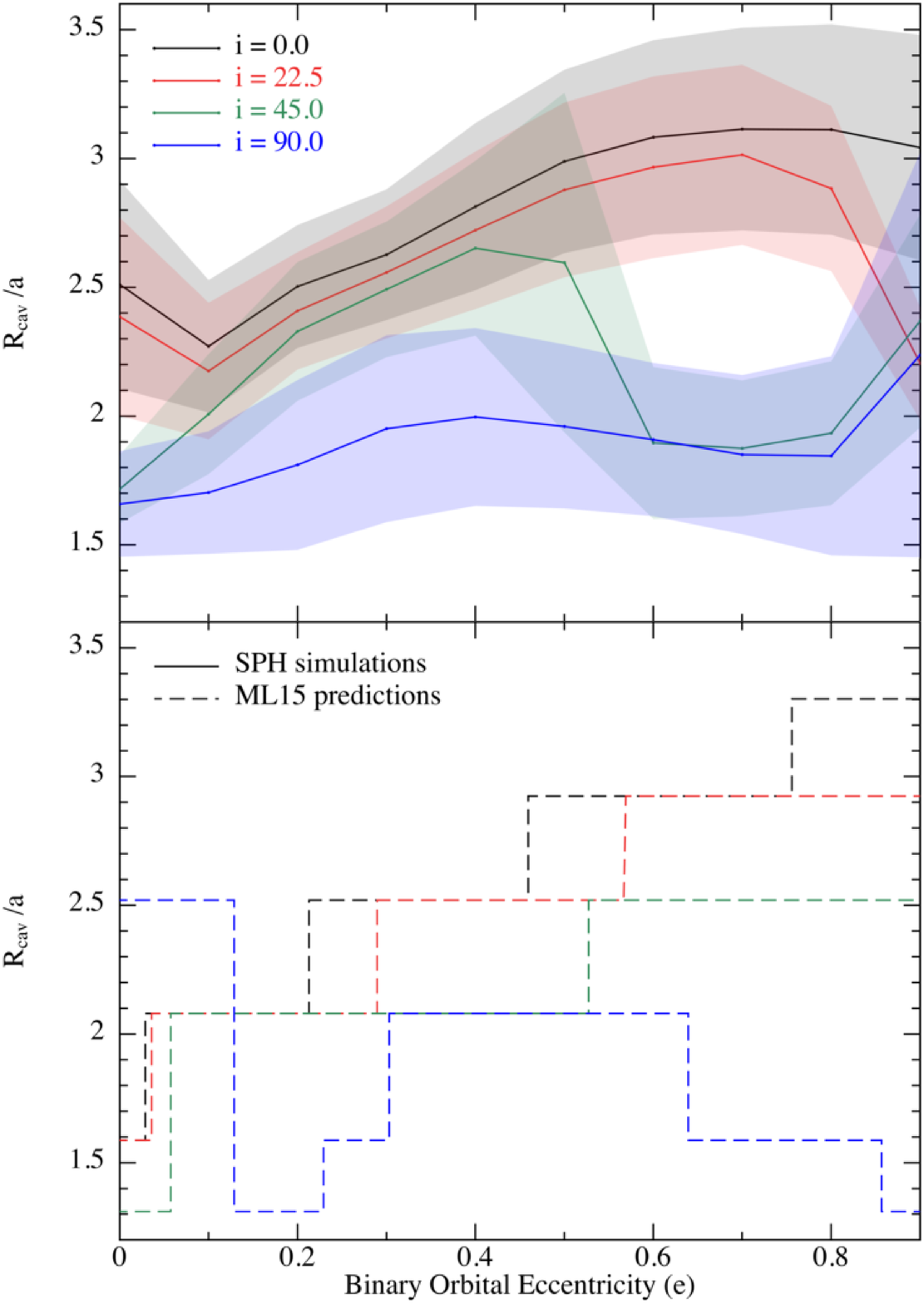}
    \caption{Cavity size as a function of binary orbital eccentricity for a disc with $(H/R)_{\rm{in}} = 0.05$ surrounding a binary with $q = 0.1$ after 1000 binary orbits. Different line colours depict discs with different initial inclinations to the binary orbital plane. The solid lines (top panel) represent the results from our SPH simulations while the dashed lines (bottom panel) represent the analytical estimates from \citetalias{ML15}.
    }
    \label{fig:cav_vs_inc}
\end{figure}

\subsection{Disc Inclination}
\label{Sec:disc_incl}

Figure~\ref{fig:q01_incl_rendered_both} shows circumbinary discs with $q = 0.1$, $(H/R)_{\rm{in}} = 0.05$ and $e = 0.8$ in both face-on and side-on views, rendered in surface density, with various initial inclinations and at various times.
The critical inclination above which a disc of test particles tends towards a polar alignment is given by \citet{Hossam18}:
\begin{equation}
    \label{eq:crit_incl}
    i_{\rm crit} = \tan^{-1}\sqrt{\frac{1 - e^2}{5e^2}}.
\end{equation}
For $e = 0.8$ this corresponds to a critical inclination of $18.5^\circ$.
Discs with an initial inclination lower than $i_{\rm crit}$ will tend towards a coplanar orbit.
If the alignment time is shorter than the lifetime of the disc, this result implies that the final configurations will always be either polar or coplanar.
In the case where the cavity is opened faster than the final alignment is reached, the disc will pass through a sequence of quasi-stationary configurations where the cavity size decreases (increases) as the disc progressively moves towards the polar (coplanar) configuration.
In the case where the final alignment is reached before the cavity is opened, the cavity will be the same size as that of a disc initially in the final configuration.
In Figure~\ref{fig:q01_incl_rendered_both} we see that, consistent with Equation~\ref{eq:crit_incl}, both the $i = 22.5^\circ$ and the $i = 45^\circ$ discs tend towards a polar alignment, but their evolution looks very different due to the different alignment times.

For the $i = 45^{\circ}$ disc the binary torque is strong enough to break the disc \citep[c.f.][]{Nixon13,Facchini13} and the inner disc quickly goes polar within hundreds of binary orbits; that is to say that the inner disc reaches a polar configuration on the same timescale as the cavity is opened. The outer disc aligns more slowly due to the weakened interaction with the binary.

For the $i = 22.5^{\circ}$ disc the binary torque is not strong enough to break the disc and instead a warp forms in the inner regions of the disc which moves outwards over time.
In this case the disc tends towards a polar alignment on the order of thousands of binary orbits while rigidly precessing.
From the third row of Figure~\ref{fig:q01_incl_rendered_both} we see that after 100 binary orbits the inner disc remains at a low inclination.
Comparing the first two rows of Figure~\ref{fig:q01_incl_rendered_both} we see the cavity opening process is similar to an initially coplanar disc due to the low inclination during the opening timescale.

Not shown in Figure~\ref{fig:q01_incl_rendered_both} are discs with an initial inclination less than $i_{\rm crit}$ ($e \leq 0.4$ for $i = 22.5^\circ$ and $e \leq 0.7$ for $i = 45^\circ$).
These discs tend towards a coplanar alignment and for $q = 0.1$ do so by warping.

The effect that the differing evolution has on the cavity size can be seen in Figure~\ref{fig:cav_vs_inc}, which shows the cavity size of inclined discs after 1000 binary orbits.
The two major factors in determining the cavity size at this time are whether the disc tends towards a coplanar or a polar alignment, and how quickly this alignment is reached.

Discs that tend towards a coplanar alignment open a cavity that is slightly smaller than that of an initially coplanar disc, due to the weaker binary torques in an inclined disc \citepalias[c.f.][]{ML15}.
The cavity then grows in time as the inclination is damped and the long term cavity size is expected to be that of an initially coplanar disc, though the realignment time is longer than the 1000 binary orbits we simulated.

As discussed earlier, discs that break and go polar reach a polar configuration within 100 binary orbits.
This means that the cavity is opened when the disc is already polar, so the cavity size is equal to that of an initially polar disc.

Discs that warp and go polar do so slowly enough to open a cavity at an intermediate inclination before their inclination starts to increase.
The binary torques get weaker as the disc gets more inclined, allowing the cavity to shrink as it is filled in due to viscous spreading.
This process requires the disc to be evolved for a viscous time at its final polar configuration before the long-term cavity size can be recovered.
After 1000 binary orbits, however, we recover an intermediate cavity size as the disc is still in the process of shrinking its cavity.

Simulations with $q = 0.5$ (not shown) produce similar results, with the exception that the binary torques are strong enough to break the disc, regardless of the binary eccentricity and disc inclination, leading to a faster alignment to either a coplanar or polar orbit.
This leads to a cavity size that is equal to that of an initially polar disc for any disc that goes polar, while discs that tend towards a coplanar alignment again have a cavity size slightly smaller than that of an initially coplanar disc.
For the coplanar discs breaking instead of warping allows for faster realignment for the coplanar discs, especially at low eccentricity. 
This leads to a cavity size that is closer to that of an initially coplanar disc.

\subsection{Binary Mass Ratio}
\label{Sec:mass_ratio}

\begin{figure*}
	\includegraphics[width=\textwidth]{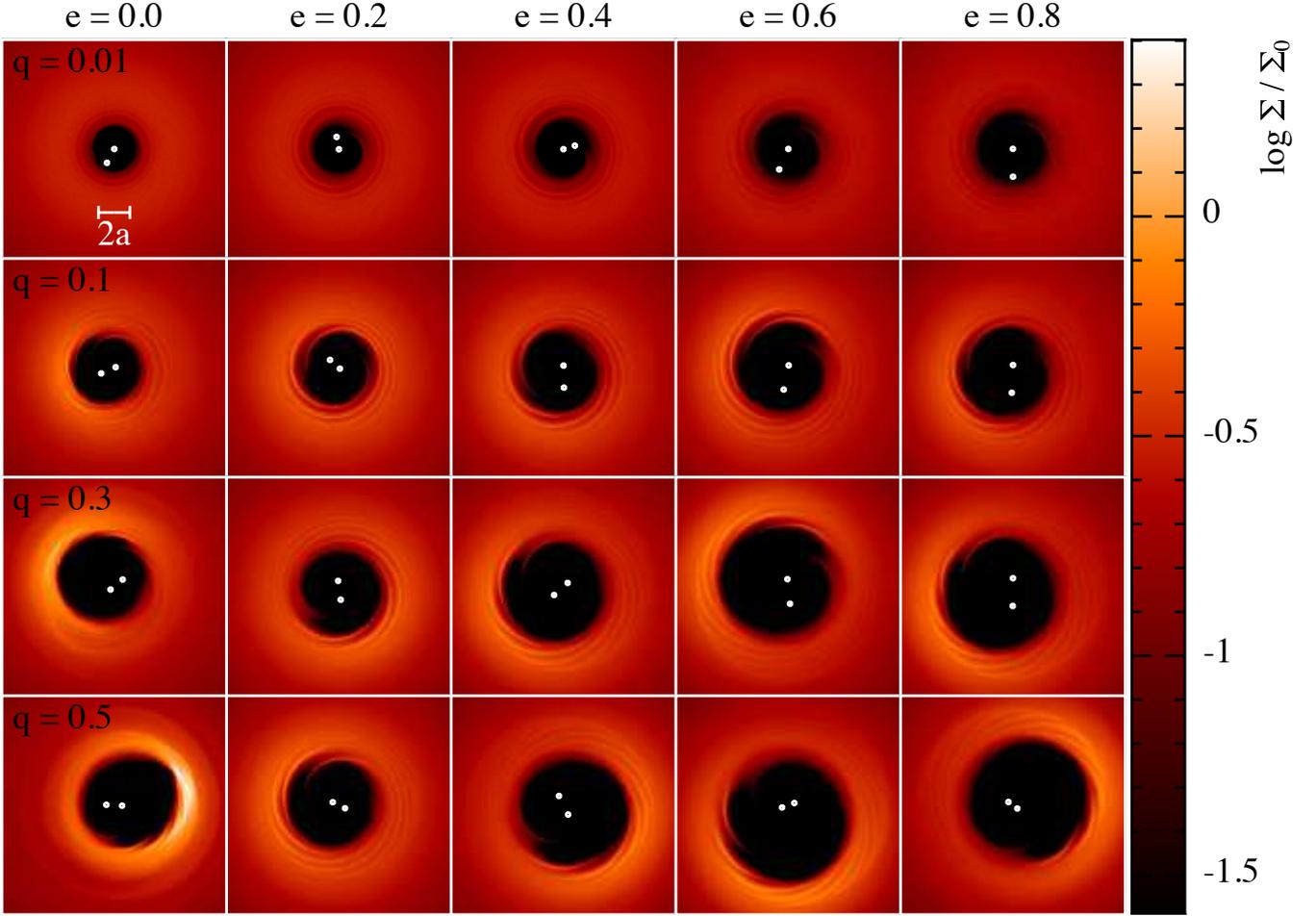}
    \caption{Surface density rendered face-on views of coplanar circumbinary discs with $(H/R)_{\rm{in}} = 0.05$ surrounding a binary with varying mass ratio and eccentricity after 1000 binary orbits. Binary mass ratio increases top to bottom, binary orbital eccentricity increases left to right. 
    }
    \label{fig:mass_ratio_rendered}
\end{figure*}

Figure~\ref{fig:mass_ratio_rendered} shows surface density rendered face-on views of circumbinary discs with $(H/R)_{\rm{in}} = 0.05$ after $1000$ binary orbits for various eccentricities and binary mass ratios.
Around circular binaries, strong horseshoe shaped over-densities are seen at the cavity edge, becoming weaker as the companion decreases in mass and disappearing at $q = 0.001$.
Faint horseshoes can also be seen around highly eccentric binaries, again becoming weaker with smaller companions.

\begin{figure}
	\includegraphics[width=0.45\textwidth]{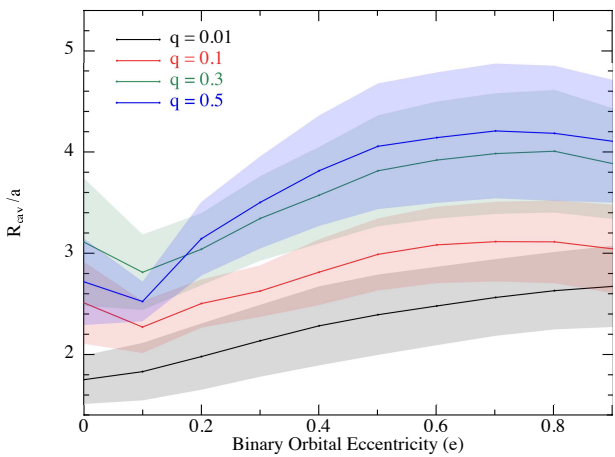}
    \caption{Cavity size as a function of binary orbital eccentricity for a coplanar disc with $(H/R)_{\rm{in}} = 0.05$ after 1000 binary orbits. Different lines depict binaries with different mass ratios. 
    }
    \label{fig:mass_ratio}
\end{figure}

Figure~\ref{fig:mass_ratio} shows the cavity size as a function of eccentricity for coplanar discs with $(H/R)_{\rm{in}} = 0.05$ around binaries with four different mass ratios ($q = 0.01, 0.1, 0.3 \text{ and } 0.5$) after 1000 binary orbits.
When $q \geq 0.1$ we see the turnover discussed in \S\ref{sec:binary_eccen}.
Consistent with \citet{Ragusa17}, we find that the more massive companions carve the largest cavities.
There is an exception to this at low eccentricity ($e \leq 0.2$) where the maximum cavity size is seen around binaries with $q = 0.3$, though we caution here that our resolution in mass ratio is coarse.

\subsection{Gas Depletion}

\begin{figure}
	\includegraphics[width=0.45\textwidth]{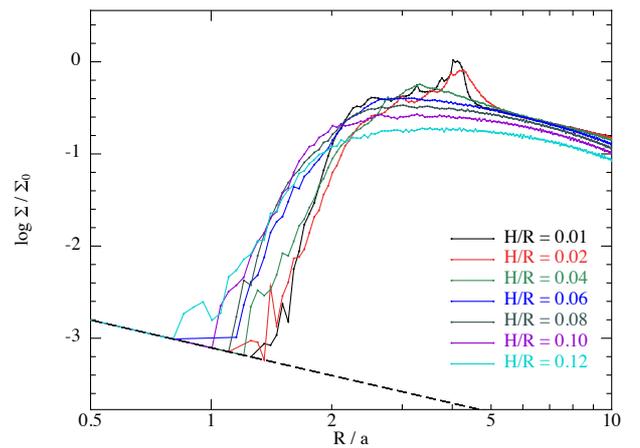}
    \caption{Azimuthally averaged surface density as a function of radius in the disc, in units of binary semi-major axis, on a log-log scale. All discs are coplanar with a circular binary of $q = 0.1$, with each line representing a different disc aspect ratio. The dashed line is the lowest possible $\Sigma$ we can resolve.
    }
    \label{fig:depletion}
\end{figure}

Figure~\ref{fig:depletion} shows the azimuthally averaged surface density as a function of radius for coplanar discs around circular binaries with $q = 0.1$ and various disc aspect ratios, allowing us to see how \removed{the} depleted the cavity is.
One common method to characterise the depletion is to model the surface density of the disc as a power law with an exponential taper \citep{LBP74} and model the cavity by scaling down the surface density by a constant depletion factor, giving \citep{Andrews11,Perez15}
\begin{equation}
    \label{eq:cavity_depletion}
    \Sigma(r) = \delta_\text{gap} \Sigma_0 \left(\frac{r}{R_0}\right)^{-p} \exp{\left[\left(-\frac{r}{R_0}\right)^{2 - p}\right]},
\end{equation}
where $\Sigma_0$ is the surface density at the characteristic radius $R_0$ and $\delta_\text{gap}$ is the depletion factor, with $\delta_\text{gap} = 1$ outside the cavity and $\delta_\text{gap} < 1$ inside the cavity.
This characterisation is impossible for us since the surface density inside the cavity is below what we are able to resolve, so instead we take the depletion as $\delta_{\rm gap} = \Sigma_{\rm max}/\Sigma_{\rm min}$, where $\Sigma_{\rm max}$ and $\Sigma_{\rm min}$ and the maximum and minimum values of surface density that we recover, respectively.

We find that in all cases the depletion is 2--3 orders of magnitude, decreasing as the disc becomes more viscous.
However, in every case our $\Sigma_{\rm min}$ is at our resolution limit, so these values can only be treated as a minimum depletion in the cavity.
Furthermore, the more viscous discs have a smaller $\delta_\text{gap}$ due to a smaller $\Sigma_{\rm max}$, which does not necessarily imply a less depleted cavity.

\section{Discussion}
\label{sec:discussion}

When comparing to previous works we found results consistent with the main conclusions from \citetalias{AL94} and \citetalias{ML15}, namely that cavity size increases with increasing binary eccentricity and decreasing disc viscosity.

\citetalias{ML15} also found that for discs with $i \leq 45^\circ$ cavity size decreases with disc inclination (bottom panel of Fig.~\ref{fig:cav_vs_inc}).
However theirs was an analytical study, comparing the strengths of the viscous and binary torques for static discs and not taking into account changes in inclination over time.
For discs which tend towards a coplanar orbit we also find that cavity size decreases with initial inclination, however for discs which tend towards a polar orbit this is no longer the case (see Section~\ref{Sec:disc_incl}).

\citet{Thun17} found cavities that extend to upwards of seven times the binary semi-major axis for coplanar discs with $(H/R)_{\rm{in}} = 0.05$ around highly eccentric binaries.
They also found a turnover in cavity size as a function of binary eccentricity that persists to 16,000 binary orbits, with the smallest cavities being opened by binaries with $e = 0.18$.
We also found a turnover (see Fig.~\ref{fig:q01_i000_HR05_cav_vs_eccen}), however ours was at $e = 0.1$ and only appears at intermediate stages of the disc evolution, disappearing after several thousands of binary orbits.
These differences are likely due to the different codes used in our analyses.
The use of 2D grid-based codes requires careful consideration of the inner boundary when working with a polar grid \citep[see Section 4.1 of ][]{Thun17}.
Indeed, recent works have found that choosing an open boundary with $R_{\rm in} > a$, as in \citet{Thun17}, can lead to an artificially large cavity \citep{Mutter17,Pierens20}.

\begin{figure}
	\includegraphics[width=0.5\textwidth]{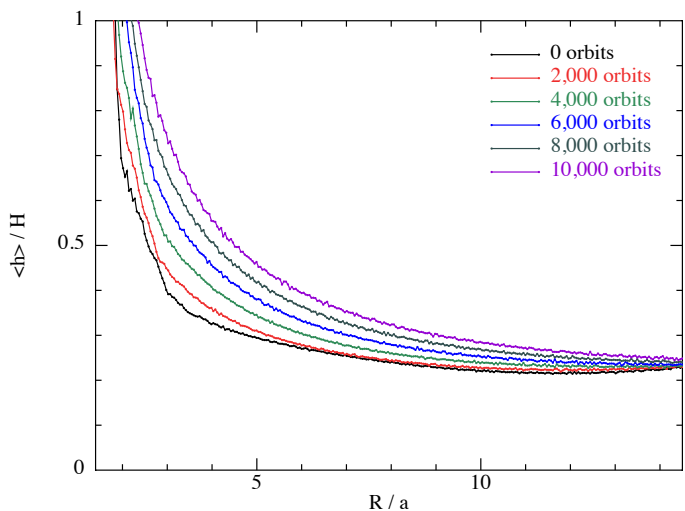}
    \caption{Smoothing length, in units of disc scale height, as a function of radius in the disc, in units of binary semi-major axis for a coplanar disc with $(H/R)_{\rm{in}} = 0.05$ around a circular binary with $q = 0.1$. Lines are shown at 20,000 binary orbit intervals.
    }
    \label{fig:q01_i000_HR05_smooth}
\end{figure}

\citet{Ragusa17} find that larger companions carve larger cavities, though they only consider binaries with $q \leq 0.2$.
As discussed in Section~\ref{Sec:mass_ratio} we found this to be the case for binaries with $q \leq 0.3$.
Equal mass binaries continue this trend when highly eccentric, but when $e \lessapprox 0.2$ equal mass binaries carve smaller cavities than those with $q = 0.3$. The reason for this is unclear.

\begin{figure}
	\includegraphics[width=0.5\textwidth]{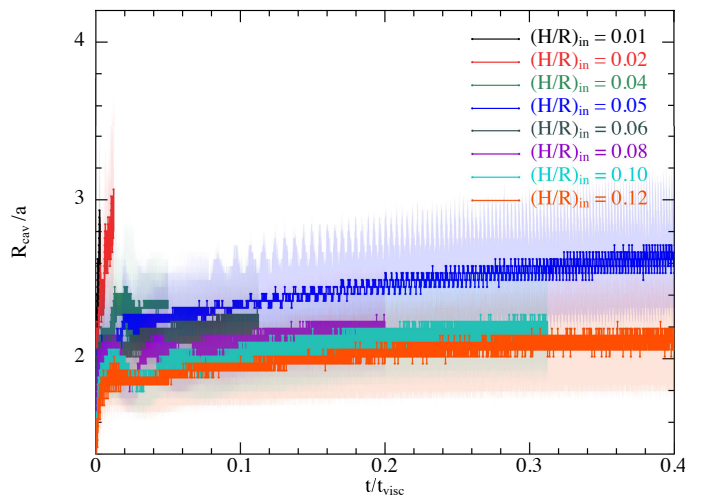}
    \caption{Cavity size as a function of fraction of viscous time for coplanar discs surrounding a binary with $e = 0.1$.
    }
    \label{fig:q01_i000_e010_cav_vs_visc_time}
\end{figure}

In Section~\ref{Sec:scale_height} we compare discs with different scale heights at the same number of orbits, corresponding to a different fraction of the viscous time for each disc.
It is natural to consider whether comparing the discs at the same viscous time shows the same behaviour.
To this end Fig.~\ref{fig:q01_i000_e010_cav_vs_visc_time} shows the evolution of the cavity size as a fraction of the viscous time.
After $\sim 10\%$ of a viscous time the cavity size is seen to increase with decreasing viscosity, as in Section~\ref{Sec:scale_height}.
That is, our conclusions are independent of whether we use time in orbits, or time in viscous times.
Figure~\ref{fig:q01_i000_e010_cav_vs_visc_time} suggests that even low eccentricity ($e=0.1$) binaries in discs with $(H/R)_{\rm in} \lesssim 0.02$ may eventually produce cavities with radii $\gtrsim 3$--4 times the semi-major axis.
However, evolving such low viscosity discs for a significant fraction of the viscous time is prohibitively expensive.
The results are also irrelevant for protoplanetary discs, since in these cases the viscous time starts to exceed the disc lifetime.

When comparing this work to observations, one should keep in mind the following:

We only modelled the gas.
ALMA continuum observations probe the mm-sized dust in the disc midplane.
While comparisons to ALMA observation of gas lines can be made with gas-only simulations, comparisons to the continuum observations would require full gas and dust simulations.
We would expect to see a larger cavity in the dust than in the gas \citep[c.f.][]{van-der-Marel18}.

We prescribed a fixed temperature profile.
Since we did not account for radiation from the central stars  we have an axisymmetric temperature profile, rather than one which oscillates during the orbit of the stars \citep{Nagel10,BQ15}.
This azimuthal temperature variation would lead to local fluctuations in viscosity.
Any effects from these local fluctuations would be minor, since they oscillate on the orbital time, while the cavity size is set on the viscous time.
The temperature difference caused by shadows from a circumstellar disc \citep[e.g. HD 142527,][]{Casassus15b,Avenhaus17} would be persistent and could drive an eccentric cavity due to local regions of low viscosity.
Furthermore, once a cavity becomes eccentric it would have a non-axisymmetric temperature profile, with the material at periastron being warmer than that at apoastron.
This could drive larger eccentricity in the cavity, though a simulation that combines dynamics and radiative transfer would be needed to investigate this effect \citep[see][]{Nealon20}. Recently, \citet{MR19b} showed that using a locally isothermal equation of state --- instead of solving the energy equation --- may overestimate the contrast of gaps in discs. Whether or not this would change our conclusions regarding the cavity size would be worthy of investigation.

We model viscosity using an $\alpha$-disc with $\alpha = 5\times10^{-3}$, though the true value of $\alpha$ in protoplanetary discs remains uncertain.
\citet{Flaherty20} argue that most observational evidence points towards $\alpha \sim 10^{-4} - 10^{-3}$ \citep[see also][]{MD12,Boneberg16}.
They do, however, provide examples of a small number of systems with $\alpha \sim 10^{-2}$, suggesting that higher values of $\alpha$, while uncommon, are possible.
Contrasting this, \citet{Papaloizou05b} found that the resonant coupling between inertial-gravity waves and a free $m = 1$ global mode causes an instability which leads to a turbulence with an effective $\alpha \sim 10^{-3}$.
\citet{Pierens20} found that the presence of a binary strengthens this instability, leading to an effective $\alpha \sim 5\times10^{-3}$.
It is also important to note that setting a low $\alpha_\text{AV}$ in SPH leads to a higher than expected dissipation \citep{MB12}.
This effect is reduced with higher resolution simulations, however simulating an $\alpha$ of $10^{-4}$ would require an unfeasibly large number of particles.
Since a higher viscosity leads to a smaller cavity, our results would underestimate the cavity size in discs with low $\alpha$.

In section~\ref{Sec:disc_incl} we claim that all discs will tend towards either a coplanar or a polar orbit in the long term.
While this is valid for low mass discs, \citet{ML19} found that discs with a significantly high mass can reach an equilibrium at an intermediate inclination.
As such, as study investigating the effects of disc mass would be required to understand how the cavity evolves in these high mass discs.

Viscous disc spreading reduces the surface density of the disc over time.
This also reduces the number density of SPH particles in the disc, lowering the resolution and increasing the smoothing length over time, as shown in Fig.~\ref{fig:q01_i000_HR05_smooth}.
Due to our prescription of disc viscosity (Section~\ref{sec:disc_visc}), this leads to an increase in the viscosity over time.
The effect is more pronounced in the inner disc, where the smoothing length increases by up to a factor of 4 after 10,000 binary orbits.
An increased viscosity near the cavity edge may lead to an underestimated cavity size in our simulations, though since this effect is present in all discs we expect the trends to remain unaffected.

In our simulations we found that cavities can be up to five times the semi-major axis of the binary orbit, meaning that binary companions can be close to the primary while still carving a large cavity, making the companion difficult to resolve.
Indeed, many discs previously classified as transitional, such as CoKu Tau/4 \citep{IK08} and HD 142527 \citep{Biller12}, have only been reclassified as circumbinary in the last 15 years, despite hosting stellar or sub-stellar mass companions.
This problem is further exacerbated for highly eccentric discs since they tend towards a polar inclination \citep{Hossam15,MarLub18}.
For any polar disc (or indeed, any highly-inclined disc) in a nearly face-on configuration, the small projected separation of the binary would make it extremely difficult to resolve.
Since highly-inclined discs are not uncommon for binaries with long periods \citep{Czekala19}, many discs currently classified as transitional may yet turn out to be hiding binary companions.

\section{Conclusions}


We have performed a numerical examination of cavity opening by a binary embedded in an accretion disc, revisiting and expanding the original numerical and analytic study by \citetalias{AL94} and the recent analytic extension to inclined discs by \citetalias{ML15}.
We considered the effects of binary eccentricity, inclination, mass ratio, disc vertical scale height and binary mass ratio.
Our conclusions are:
\begin{enumerate}
\item There exists two timescales for the cavity opening process.
The cavity is quickly opened on a dynamical time, within a few tens of orbits, while the long term size of the cavity is set on the viscous time, after tens of thousands of orbits.
\item Binaries carve cavities in circumbinary discs at a radius 2--5 times the semi-major axis, with a cavity size that depends most strongly on binary eccentricity and disc viscosity, as predicted by \citetalias{AL94} and \citetalias{ML15}.
\item When considering inclined discs there exists three regimes.
Discs which evolve towards a coplanar orbit have a cavity size slightly smaller than an initially coplanar disc, decreasing in size as initial inclination increases.
Discs which evolve towards a polar orbit have a cavity size which depends on their evolution, i.e. whether they break or warp.
Discs which break have a small cavity equal in size to those of an initially polar disc, while discs that warp quickly open a large cavity which is then filled in on the viscous timescale, resulting in an intermediate sized cavity which shrinks on a viscous timescale to the size of a cavity in an initially polar disc.
\item Cavity size is an increasing function of binary mass ratio for all but the largest companions on low eccentricity orbits.
\item All of our binaries (with $q \geq 0.01$) produced a gas depletion inside the cavity of at least 2 orders of magnitude in surface density.

\end{enumerate}

\section*{Acknowledgements}
We thank the anonymous referee for their useful comments which improved the manuscript.
We thank Giuseppe Lodato, Hossam Aly and Arnaud Vericel for stimulating discussions.
We acknowledge use of the Ozstar supercomputer, funded by the Australian Government and Swinburne University, and use of Magnus and Raijin via the Australian National Compute Infrastructure.
We acknowledge funding from the European Union's H2020 research and innovation programme under Marie Sk\l{}odowska-Curie grant agreement No 823823.
KH and JFG acknowledge funding from ANR (Agence Nationale de la Recherche) of France via ANR-16-CE31-0013 (Planet-Forming-Disks) and thank the LABEX Lyon Institute of Origins (ANR-10-LABX-0066) of the Universit\'e de Lyon for financial support within the programme `Investissements d'Avenir' (ANR-11-IDEX-0007) of the French government operated by the ANR.
DJP acknowledges funding from the Australian Research Council under FT130100034 and DP180104235. ER acknowledges financial support from the European Research Council (ERC) under the European Union's Horizon 2020 research and innovation programme (grant agreement No 681601).
Plots were made with \textsc{splash} \citep{splash}.

\section*{Data Availability Statement}

The \textsc{Phantom} SPH code is available at \url{https://github.com/danieljprice/phantom}. The input files for generating our SPH simulations are available upon request.





\bibliographystyle{mnras}
\bibliography{Kieran} 

\begin{thebibliography}{}
\makeatletter
\relax
\def\mn@urlcharsother{\let\do\@makeother \do\$\do\&\do\#\do\^\do\_\do\%\do\~}
\def\mn@doi{\begingroup\mn@urlcharsother \@ifnextchar [ {\mn@doi@}
  {\mn@doi@[]}}
\def\mn@doi@[#1]#2{\def\@tempa{#1}\ifx\@tempa\@empty \href
  {http://dx.doi.org/#2} {doi:#2}\else \href {http://dx.doi.org/#2} {#1}\fi
  \endgroup}
\def\mn@eprint#1#2{\mn@eprint@#1:#2::\@nil}
\def\mn@eprint@arXiv#1{\href {http://arxiv.org/abs/#1} {{\tt arXiv:#1}}}
\def\mn@eprint@dblp#1{\href {http://dblp.uni-trier.de/rec/bibtex/#1.xml}
  {dblp:#1}}
\def\mn@eprint@#1:#2:#3:#4\@nil{\def\@tempa {#1}\def\@tempb {#2}\def\@tempc
  {#3}\ifx \@tempc \@empty \let \@tempc \@tempb \let \@tempb \@tempa \fi \ifx
  \@tempb \@empty \def\@tempb {arXiv}\fi \@ifundefined
  {mn@eprint@\@tempb}{\@tempb:\@tempc}{\expandafter \expandafter \csname
  mn@eprint@\@tempb\endcsname \expandafter{\@tempc}}}

\bibitem[\protect\citeauthoryear{{Alexander} \& {Armitage}}{{Alexander} \&
  {Armitage}}{2007}]{AA07}
{Alexander} R.~D.,  {Armitage} P.~J.,  2007, \mn@doi [\mnras]
  {10.1111/j.1365-2966.2006.11341.x}, \href
  {https://ui.adsabs.harvard.edu/abs/2007MNRAS.375..500A} {375, 500}

\bibitem[\protect\citeauthoryear{{Aly}, {Dehnen}, {Nixon}  \& {King}}{{Aly}
  et~al.}{2015}]{Hossam15}
{Aly} H.,  {Dehnen} W.,  {Nixon} C.,   {King} A.,  2015, \mn@doi [\mnras]
  {10.1093/mnras/stv128}, \href
  {http://adsabs.harvard.edu/abs/2015MNRAS.449...65A} {449, 65}

\bibitem[\protect\citeauthoryear{{Aly}, {Lodato}  \& {Cazzoletti}}{{Aly}
  et~al.}{2018}]{Hossam18}
{Aly} H.,  {Lodato} G.,   {Cazzoletti} P.,  2018, \mn@doi [\mnras]
  {10.1093/mnras/sty2179}, \href
  {https://ui.adsabs.harvard.edu/abs/2018MNRAS.480.4738A} {480, 4738}

\bibitem[\protect\citeauthoryear{{Andrews}, {Wilner}, {Espaillat}, {Hughes},
  {Dullemond}, {McClure}, {Qi}  \& {Brown}}{{Andrews} et~al.}{2011}]{Andrews11}
{Andrews} S.~M.,  {Wilner} D.~J.,  {Espaillat} C.,  {Hughes} A.~M.,
  {Dullemond} C.~P.,  {McClure} M.~K.,  {Qi} C.,   {Brown} J.~M.,  2011,
  \mn@doi [\apj] {10.1088/0004-637X/732/1/42}, \href
  {http://adsabs.harvard.edu/abs/2011ApJ...732...42A} {732, 42}

\bibitem[\protect\citeauthoryear{{Artymowicz} \& {Lubow}}{{Artymowicz} \&
  {Lubow}}{1994}]{AL94}
{Artymowicz} P.,  {Lubow} S.~H.,  1994, \mn@doi [\apj] {10.1086/173679}, \href
  {http://adsabs.harvard.edu/abs/1994ApJ...421..651A} {421, 651}

\bibitem[\protect\citeauthoryear{{Ataiee}, {Pinilla}, {Zsom}, {Dullemond},
  {Dominik}  \& {Ghanbari}}{{Ataiee} et~al.}{2013}]{Ataiee13}
{Ataiee} S.,  {Pinilla} P.,  {Zsom} A.,  {Dullemond} C.~P.,  {Dominik} C.,
  {Ghanbari} J.,  2013, \mn@doi [\aap] {10.1051/0004-6361/201321125}, \href
  {https://ui.adsabs.harvard.edu/abs/2013A&A...553L...3A} {553, L3}

\bibitem[\protect\citeauthoryear{{Avenhaus} et~al.,}{{Avenhaus}
  et~al.}{2017}]{Avenhaus17}
{Avenhaus} H.,  et~al., 2017, \mn@doi [\aj] {10.3847/1538-3881/aa7560}, \href
  {http://adsabs.harvard.edu/abs/2017AJ....154...33A} {154, 33}

\bibitem[\protect\citeauthoryear{{Bate}, {Bonnell}  \& {Price}}{{Bate}
  et~al.}{1995}]{Bate95}
{Bate} M.~R.,  {Bonnell} I.~A.,   {Price} N.~M.,  1995, \mn@doi [\mnras]
  {10.1093/mnras/277.2.362}, \href
  {http://adsabs.harvard.edu/abs/1995MNRAS.277..362B} {277, 362}

\bibitem[\protect\citeauthoryear{{Benisty} et~al.,}{{Benisty}
  et~al.}{2017}]{Benisty17}
{Benisty} M.,  et~al., 2017, \mn@doi [\aap] {10.1051/0004-6361/201629798},
  \href {http://adsabs.harvard.edu/abs/2017A%26A...597A..42B} {597, A42}

\bibitem[\protect\citeauthoryear{{Biller} et~al.,}{{Biller}
  et~al.}{2012}]{Biller12}
{Biller} B.,  et~al., 2012, \mn@doi [\apjl] {10.1088/2041-8205/753/2/L38},
  \href {https://ui.adsabs.harvard.edu/abs/2012ApJ...753L..38B} {753, L38}

\bibitem[\protect\citeauthoryear{{Bodman} \& {Quillen}}{{Bodman} \&
  {Quillen}}{2015}]{BQ15}
{Bodman} E. H.~L.,  {Quillen} A.,  2015, \mn@doi [\mnras]
  {10.1093/mnras/stv1769}, \href
  {https://ui.adsabs.harvard.edu/abs/2015MNRAS.453.2387B} {453, 2387}

\bibitem[\protect\citeauthoryear{{Boneberg}, {Pani{\'c}}, {Haworth}, {Clarke}
  \& {Min}}{{Boneberg} et~al.}{2016}]{Boneberg16}
{Boneberg} D.~M.,  {Pani{\'c}} O.,  {Haworth} T.~J.,  {Clarke} C.~J.,   {Min}
  M.,  2016, \mn@doi [\mnras] {10.1093/mnras/stw1325}, \href
  {https://ui.adsabs.harvard.edu/abs/2016MNRAS.461..385B} {461, 385}

\bibitem[\protect\citeauthoryear{{Brice{\~n}o} \& {Tokovinin}}{{Brice{\~n}o} \&
  {Tokovinin}}{2017}]{Briceno17}
{Brice{\~n}o} C.,  {Tokovinin} A.,  2017, \mn@doi [\aj]
  {10.3847/1538-3881/aa8e9b}, \href
  {https://ui.adsabs.harvard.edu/abs/2017AJ....154..195B} {154, 195}

\bibitem[\protect\citeauthoryear{{Calcino}, {Price}, {Pinte}, {van der Marel},
  {Ragusa}, {Dipierro}, {Cuello}  \& {Christiaens}}{{Calcino}
  et~al.}{2019}]{Calcino19}
{Calcino} J.,  {Price} D.~J.,  {Pinte} C.,  {van der Marel} N.,  {Ragusa} E.,
  {Dipierro} G.,  {Cuello} N.,   {Christiaens} V.,  2019, \mn@doi [\mnras]
  {10.1093/mnras/stz2770}, \href
  {https://ui.adsabs.harvard.edu/abs/2019MNRAS.490.2579C} {490, 2579}

\bibitem[\protect\citeauthoryear{{Canovas} et~al.,}{{Canovas}
  et~al.}{2018}]{Canovas18}
{Canovas} H.,  et~al., 2018, \mn@doi [\aap] {10.1051/0004-6361/201731640},
  \href {https://ui.adsabs.harvard.edu/abs/2018A%26A...610A..13C} {610, A13}

\bibitem[\protect\citeauthoryear{{Casassus} et~al.,}{{Casassus}
  et~al.}{2013}]{Casassus13}
{Casassus} S.,  et~al., 2013, \mn@doi [\nat] {10.1038/nature11769}, \href
  {http://adsabs.harvard.edu/abs/2013Natur.493..191C} {493, 191}

\bibitem[\protect\citeauthoryear{{Casassus} et~al.,}{{Casassus}
  et~al.}{2015}]{Casassus15b}
{Casassus} S.,  et~al., 2015, \mn@doi [\apj] {10.1088/0004-637X/812/2/126},
  \href {http://adsabs.harvard.edu/abs/2015ApJ...812..126C} {812, 126}

\bibitem[\protect\citeauthoryear{{Casassus} et~al.,}{{Casassus}
  et~al.}{2018}]{Casassus18}
{Casassus} S.,  et~al., 2018, \mn@doi [\mnras] {10.1093/mnras/sty894}, \href
  {http://adsabs.harvard.edu/abs/2018MNRAS.477.5104C} {477, 5104}

\bibitem[\protect\citeauthoryear{{Cazzoletti}, {Ricci}, {Birnstiel}  \&
  {Lodato}}{{Cazzoletti} et~al.}{2017}]{Cazzoletti17}
{Cazzoletti} P.,  {Ricci} L.,  {Birnstiel} T.,   {Lodato} G.,  2017, \mn@doi
  [\aap] {10.1051/0004-6361/201629721}, \href
  {http://adsabs.harvard.edu/abs/2017A%26A...599A.102C} {599, A102}

\bibitem[\protect\citeauthoryear{{Clarke}, {Gendrin}  \& {Sotomayor}}{{Clarke}
  et~al.}{2001}]{Clarke01}
{Clarke} C.~J.,  {Gendrin} A.,   {Sotomayor} M.,  2001, \mn@doi [\mnras]
  {10.1046/j.1365-8711.2001.04891.x}, \href
  {https://ui.adsabs.harvard.edu/abs/2001MNRAS.328..485C} {328, 485}

\bibitem[\protect\citeauthoryear{{Czekala}, {Chiang}, {Andrews}, {Jensen},
  {Torres}, {Wilner}, {Stassun}  \& {Macintosh}}{{Czekala}
  et~al.}{2019}]{Czekala19}
{Czekala} I.,  {Chiang} E.,  {Andrews} S.~M.,  {Jensen} E.~L.~N.,  {Torres} G.,
   {Wilner} D.~J.,  {Stassun} K.~G.,   {Macintosh} B.,  2019, arXiv e-prints,
  \href {https://ui.adsabs.harvard.edu/abs/2019arXiv190603269C} {}

\bibitem[\protect\citeauthoryear{{Dong}, {Zhu}, {Rafikov}  \& {Stone}}{{Dong}
  et~al.}{2015}]{Dong15a}
{Dong} R.,  {Zhu} Z.,  {Rafikov} R.~R.,   {Stone} J.~M.,  2015, \mn@doi [\apjl]
  {10.1088/2041-8205/809/1/L5}, \href
  {https://ui.adsabs.harvard.edu/abs/2015ApJ...809L...5D} {809, L5}

\bibitem[\protect\citeauthoryear{{Facchini}, {Lodato}  \& {Price}}{{Facchini}
  et~al.}{2013}]{Facchini13}
{Facchini} S.,  {Lodato} G.,   {Price} D.~J.,  2013, \mn@doi [\mnras]
  {10.1093/mnras/stt877}, \href
  {http://adsabs.harvard.edu/abs/2013MNRAS.433.2142F} {433, 2142}

\bibitem[\protect\citeauthoryear{{Flaherty} et~al.,}{{Flaherty}
  et~al.}{2020}]{Flaherty20}
{Flaherty} K.,  et~al., 2020, \mn@doi [\apj] {10.3847/1538-4357/ab8cc5}, \href
  {https://ui.adsabs.harvard.edu/abs/2020ApJ...895..109F} {895, 109}

\bibitem[\protect\citeauthoryear{{Goldreich} \& {Tremaine}}{{Goldreich} \&
  {Tremaine}}{1978}]{GT78}
{Goldreich} P.,  {Tremaine} S.,  1978, \mn@doi [\apj] {10.1086/156203}, \href
  {http://adsabs.harvard.edu/abs/1978ApJ...222..850G} {222, 850}

\bibitem[\protect\citeauthoryear{{Guilloteau}, {Dutrey}  \&
  {Simon}}{{Guilloteau} et~al.}{1999}]{Guilloteau99}
{Guilloteau} S.,  {Dutrey} A.,   {Simon} M.,  1999, \aap, \href
  {http://adsabs.harvard.edu/abs/1999A%26A...348..570G} {348, 570}

\bibitem[\protect\citeauthoryear{{G{\"u}nther} \& {Kley}}{{G{\"u}nther} \&
  {Kley}}{2002}]{GK02}
{G{\"u}nther} R.,  {Kley} W.,  2002, \mn@doi [\aap]
  {10.1051/0004-6361:20020407}, \href
  {https://ui.adsabs.harvard.edu/abs/2002A&A...387..550G} {387, 550}

\bibitem[\protect\citeauthoryear{{Ireland} \& {Kraus}}{{Ireland} \&
  {Kraus}}{2008}]{IK08}
{Ireland} M.~J.,  {Kraus} A.~L.,  2008, \mn@doi [\apjl] {10.1086/588216}, \href
  {https://ui.adsabs.harvard.edu/abs/2008ApJ...678L..59I} {678, L59}

\bibitem[\protect\citeauthoryear{{Lodato} \& {Price}}{{Lodato} \&
  {Price}}{2010}]{LP10}
{Lodato} G.,  {Price} D.~J.,  2010, \mn@doi [\mnras]
  {10.1111/j.1365-2966.2010.16526.x}, \href
  {http://adsabs.harvard.edu/abs/2010MNRAS.405.1212L} {405, 1212}

\bibitem[\protect\citeauthoryear{{Lynden-Bell} \& {Pringle}}{{Lynden-Bell} \&
  {Pringle}}{1974}]{LBP74}
{Lynden-Bell} D.,  {Pringle} J.~E.,  1974, \mn@doi [\mnras]
  {10.1093/mnras/168.3.603}, \href
  {http://adsabs.harvard.edu/abs/1974MNRAS.168..603L} {168, 603}

\bibitem[\protect\citeauthoryear{{Marino}, {Perez}  \& {Casassus}}{{Marino}
  et~al.}{2015}]{Marino15}
{Marino} S.,  {Perez} S.,   {Casassus} S.,  2015, \mn@doi [\apjl]
  {10.1088/2041-8205/798/2/L44}, \href
  {https://ui.adsabs.harvard.edu/abs/2015ApJ...798L..44M} {798, L44}

\bibitem[\protect\citeauthoryear{{Martin} \& {Lubow}}{{Martin} \&
  {Lubow}}{2017}]{ML17}
{Martin} R.~G.,  {Lubow} S.~H.,  2017, \mn@doi [\apjl]
  {10.3847/2041-8213/835/2/L28}, \href
  {http://adsabs.harvard.edu/abs/2017ApJ...835L..28M} {835, L28}

\bibitem[\protect\citeauthoryear{{Martin} \& {Lubow}}{{Martin} \&
  {Lubow}}{2018}]{MarLub18}
{Martin} R.~G.,  {Lubow} S.~H.,  2018, \mn@doi [\mnras]
  {10.1093/mnras/sty1648}, \href
  {https://ui.adsabs.harvard.edu/abs/2018MNRAS.479.1297M} {479, 1297}

\bibitem[\protect\citeauthoryear{{Martin} \& {Lubow}}{{Martin} \&
  {Lubow}}{2019}]{ML19}
{Martin} R.~G.,  {Lubow} S.~H.,  2019, \mn@doi [\mnras]
  {10.1093/mnras/stz2670}, \href
  {https://ui.adsabs.harvard.edu/abs/2019MNRAS.tmp.2294M} {p.~2294}

\bibitem[\protect\citeauthoryear{{Meru} \& {Bate}}{{Meru} \&
  {Bate}}{2012}]{MB12}
{Meru} F.,  {Bate} M.~R.,  2012, \mn@doi [\mnras]
  {10.1111/j.1365-2966.2012.22035.x}, \href
  {https://ui.adsabs.harvard.edu/abs/2012MNRAS.427.2022M} {427, 2022}

\bibitem[\protect\citeauthoryear{{Min}, {Stolker}, {Dominik}  \&
  {Benisty}}{{Min} et~al.}{2017}]{Min17}
{Min} M.,  {Stolker} T.,  {Dominik} C.,   {Benisty} M.,  2017, \mn@doi [\aap]
  {10.1051/0004-6361/201730949}, \href
  {http://adsabs.harvard.edu/abs/2017A%26A...604L..10M} {604, L10}

\bibitem[\protect\citeauthoryear{{Miranda} \& {Lai}}{{Miranda} \&
  {Lai}}{2015}]{ML15}
{Miranda} R.,  {Lai} D.,  2015, \mn@doi [\mnras] {10.1093/mnras/stv1450}, \href
  {http://adsabs.harvard.edu/abs/2015MNRAS.452.2396M} {452, 2396}

\bibitem[\protect\citeauthoryear{{Miranda} \& {Rafikov}}{{Miranda} \&
  {Rafikov}}{2019}]{MR19b}
{Miranda} R.,  {Rafikov} R.~R.,  2019, \mn@doi [\apjl]
  {10.3847/2041-8213/ab22a7}, \href
  {https://ui.adsabs.harvard.edu/abs/2019ApJ...878L...9M} {878, L9}

\bibitem[\protect\citeauthoryear{{Mulders} \& {Dominik}}{{Mulders} \&
  {Dominik}}{2012}]{MD12}
{Mulders} G.~D.,  {Dominik} C.,  2012, \mn@doi [\aap]
  {10.1051/0004-6361/201118127}, \href
  {https://ui.adsabs.harvard.edu/abs/2012A&A...539A...9M} {539, A9}

\bibitem[\protect\citeauthoryear{{Mutter}, {Pierens}  \& {Nelson}}{{Mutter}
  et~al.}{2017}]{Mutter17}
{Mutter} M.~M.,  {Pierens} A.,   {Nelson} R.~P.,  2017, \mn@doi [\mnras]
  {10.1093/mnras/stw2768}, \href
  {https://ui.adsabs.harvard.edu/abs/2017MNRAS.465.4735M} {465, 4735}

\bibitem[\protect\citeauthoryear{{Nagel}, {D'Alessio}, {Calvet}, {Espaillat},
  {Sargent}, {Hern{\'a}ndez}  \& {Forrest}}{{Nagel} et~al.}{2010}]{Nagel10}
{Nagel} E.,  {D'Alessio} P.,  {Calvet} N.,  {Espaillat} C.,  {Sargent} B.,
  {Hern{\'a}ndez} J.,   {Forrest} W.~J.,  2010, \mn@doi [\apj]
  {10.1088/0004-637X/708/1/38}, \href
  {https://ui.adsabs.harvard.edu/abs/2010ApJ...708...38N} {708, 38}

\bibitem[\protect\citeauthoryear{{Nealon}, {Cuello}  \& {Alexander}}{{Nealon}
  et~al.}{2020}]{Nealon20}
{Nealon} R.,  {Cuello} N.,   {Alexander} R.,  2020, \mn@doi [\mnras]
  {10.1093/mnras/stz3186}, \href
  {https://ui.adsabs.harvard.edu/abs/2020MNRAS.491.4108N} {491, 4108}

\bibitem[\protect\citeauthoryear{{Nixon} \& {Lubow}}{{Nixon} \&
  {Lubow}}{2015}]{NL15}
{Nixon} C.,  {Lubow} S.~H.,  2015, \mn@doi [\mnras] {10.1093/mnras/stv166},
  \href {https://ui.adsabs.harvard.edu/abs/2015MNRAS.448.3472N} {448, 3472}

\bibitem[\protect\citeauthoryear{{Nixon}, {King}  \& {Price}}{{Nixon}
  et~al.}{2013}]{Nixon13}
{Nixon} C.,  {King} A.,   {Price} D.,  2013, \mn@doi [\mnras]
  {10.1093/mnras/stt1136}, \href
  {https://ui.adsabs.harvard.edu/abs/2013MNRAS.434.1946N} {434, 1946}

\bibitem[\protect\citeauthoryear{{Ogilvie} \& {Lubow}}{{Ogilvie} \&
  {Lubow}}{2002}]{OL02}
{Ogilvie} G.~I.,  {Lubow} S.~H.,  2002, \mn@doi [\mnras]
  {10.1046/j.1365-8711.2002.05148.x}, \href
  {https://ui.adsabs.harvard.edu/abs/2002MNRAS.330..950O} {330, 950}

\bibitem[\protect\citeauthoryear{{Papaloizou}}{{Papaloizou}}{2005}]{Papaloizou05b}
{Papaloizou} J.~C.~B.,  2005, \mn@doi [\aap] {10.1051/0004-6361:20041948},
  \href {https://ui.adsabs.harvard.edu/abs/2005A&A...432..757P} {432, 757}

\bibitem[\protect\citeauthoryear{{Perez} et~al.,}{{Perez}
  et~al.}{2015}]{Perez15}
{Perez} S.,  et~al., 2015, \mn@doi [\apj] {10.1088/0004-637X/798/2/85}, \href
  {https://ui.adsabs.harvard.edu/abs/2015ApJ...798...85P} {798, 85}

\bibitem[\protect\citeauthoryear{{Pierens}, {McNally}  \& {Nelson}}{{Pierens}
  et~al.}{2020}]{Pierens20}
{Pierens} A.,  {McNally} C.~P.,   {Nelson} R.~P.,  2020, \mn@doi [\mnras]
  {10.1093/mnras/staa1550}, \href
  {https://ui.adsabs.harvard.edu/abs/2020MNRAS.tmp.1701P} {}

\bibitem[\protect\citeauthoryear{{Pinilla} et~al.,}{{Pinilla}
  et~al.}{2018}]{Pinilla18}
{Pinilla} P.,  et~al., 2018, \mn@doi [\apj] {10.3847/1538-4357/aabf94}, \href
  {https://ui.adsabs.harvard.edu/abs/2018ApJ...859...32P} {859, 32}

\bibitem[\protect\citeauthoryear{{Poblete}, {Calcino}, {Cuello}, {Mac{\'\i}as},
  {Ribas}, {Price}, {Cuadra}  \& {Pinte}}{{Poblete} et~al.}{2020}]{Poblete20}
{Poblete} P.~P.,  {Calcino} J.,  {Cuello} N.,  {Mac{\'\i}as} E.,  {Ribas}
  {\'A}.,  {Price} D.~J.,  {Cuadra} J.,   {Pinte} C.,  2020, \mn@doi [\mnras]
  {10.1093/mnras/staa1655}, \href
  {https://ui.adsabs.harvard.edu/abs/2020MNRAS.496.2362P} {496, 2362}

\bibitem[\protect\citeauthoryear{{Price}}{{Price}}{2007}]{splash}
{Price} D.~J.,  2007, \mn@doi [\pasa] {10.1071/AS07022}, \href
  {http://adsabs.harvard.edu/abs/2007PASA...24..159P} {24, 159}

\bibitem[\protect\citeauthoryear{{Price} et~al.,}{{Price}
  et~al.}{2018a}]{PHANTOM}
{Price} D.~J.,  et~al., 2018a, \mn@doi [\pasa] {10.1017/pasa.2018.25}, \href
  {http://adsabs.harvard.edu/abs/2018PASA...35...31P} {35, e031}

\bibitem[\protect\citeauthoryear{{Price} et~al.,}{{Price}
  et~al.}{2018b}]{Price18}
{Price} D.~J.,  et~al., 2018b, \mn@doi [\mnras] {10.1093/mnras/sty647}, \href
  {http://adsabs.harvard.edu/abs/2018MNRAS.477.1270P} {477, 1270}

\bibitem[\protect\citeauthoryear{{Pringle}}{{Pringle}}{1981}]{Pringle81}
{Pringle} J.~E.,  1981, \mn@doi [\araa] {10.1146/annurev.aa.19.090181.001033},
  \href {http://adsabs.harvard.edu/abs/1981ARA%26A..19..137P} {19, 137}

\bibitem[\protect\citeauthoryear{{Ragusa}, {Dipierro}, {Lodato}, {Laibe}  \&
  {Price}}{{Ragusa} et~al.}{2017}]{Ragusa17}
{Ragusa} E.,  {Dipierro} G.,  {Lodato} G.,  {Laibe} G.,   {Price} D.~J.,  2017,
  \mn@doi [\mnras] {10.1093/mnras/stw2456}, \href
  {http://adsabs.harvard.edu/abs/2017MNRAS.464.1449R} {464, 1449}

\bibitem[\protect\citeauthoryear{{Shakura} \& {Sunyaev}}{{Shakura} \&
  {Sunyaev}}{1973}]{SS73}
{Shakura} N.~I.,  {Sunyaev} R.~A.,  1973, \aap, \href
  {http://adsabs.harvard.edu/abs/1973A%26A....24..337S} {24, 337}

\bibitem[\protect\citeauthoryear{{Tang} et~al.,}{{Tang} et~al.}{2016}]{Tang16}
{Tang} Y.-W.,  et~al., 2016, \mn@doi [\apj] {10.3847/0004-637X/820/1/19}, \href
  {http://adsabs.harvard.edu/abs/2016ApJ...820...19T} {820, 19}

\bibitem[\protect\citeauthoryear{{Thun}, {Kley}  \& {Picogna}}{{Thun}
  et~al.}{2017}]{Thun17}
{Thun} D.,  {Kley} W.,   {Picogna} G.,  2017, \mn@doi [\aap]
  {10.1051/0004-6361/201730666}, \href
  {http://adsabs.harvard.edu/abs/2017A%26A...604A.102T} {604, A102}

\bibitem[\protect\citeauthoryear{{Tripathi}, {Andrews}, {Birnstiel}  \&
  {Wilner}}{{Tripathi} et~al.}{2017}]{Tripathi17}
{Tripathi} A.,  {Andrews} S.~M.,  {Birnstiel} T.,   {Wilner} D.~J.,  2017,
  \mn@doi [\apj] {10.3847/1538-4357/aa7c62}, \href
  {https://ui.adsabs.harvard.edu/abs/2017ApJ...845...44T} {845, 44}

\bibitem[\protect\citeauthoryear{{Tuthill}, {Monnier}  \& {Danchi}}{{Tuthill}
  et~al.}{2001}]{Tuthill01}
{Tuthill} P.~G.,  {Monnier} J.~D.,   {Danchi} W.~C.,  2001, \mn@doi [\nat]
  {10.1038/35059014}, \href
  {https://ui.adsabs.harvard.edu/abs/2001Natur.409.1012T} {409, 1012}

\bibitem[\protect\citeauthoryear{{Tuthill}, {Monnier}, {Danchi}, {Hale}  \&
  {Townes}}{{Tuthill} et~al.}{2002}]{Tuthill02}
{Tuthill} P.~G.,  {Monnier} J.~D.,  {Danchi} W.~C.,  {Hale} D.~D.~S.,
  {Townes} C.~H.,  2002, \mn@doi [\apj] {10.1086/342235}, \href
  {https://ui.adsabs.harvard.edu/abs/2002ApJ...577..826T} {577, 826}

\bibitem[\protect\citeauthoryear{{Ubeira Gabellini} et~al.,}{{Ubeira Gabellini}
  et~al.}{2019}]{Margi19}
{Ubeira Gabellini} M.~G.,  et~al., 2019, \mn@doi [\mnras]
  {10.1093/mnras/stz1138}, \href
  {https://ui.adsabs.harvard.edu/abs/2019MNRAS.486.4638U} {486, 4638}

\bibitem[\protect\citeauthoryear{{Yang} et~al.,}{{Yang} et~al.}{2017}]{Yang17}
{Yang} Y.,  et~al., 2017, \mn@doi [\aj] {10.3847/1538-3881/153/1/7}, \href
  {http://adsabs.harvard.edu/abs/2017AJ....153....7Y} {153, 7}

\bibitem[\protect\citeauthoryear{{van der Marel} et~al.,}{{van der Marel}
  et~al.}{2013}]{van-der-Marel13}
{van der Marel} N.,  et~al., 2013, \mn@doi [Science] {10.1126/science.1236770},
  \href {https://ui.adsabs.harvard.edu/abs/2013Sci...340.1199V} {340, 1199}

\bibitem[\protect\citeauthoryear{{van der Marel}, {van Dishoeck}, {Bruderer},
  {P{\'e}rez}  \& {Isella}}{{van der Marel} et~al.}{2015}]{van-der-Marel15a}
{van der Marel} N.,  {van Dishoeck} E.~F.,  {Bruderer} S.,  {P{\'e}rez} L.,
  {Isella} A.,  2015, \mn@doi [\aap] {10.1051/0004-6361/201525658}, \href
  {https://ui.adsabs.harvard.edu/abs/2015A&A...579A.106V} {579, A106}

\bibitem[\protect\citeauthoryear{{van der Marel}, {van Dishoeck}, {Bruderer},
  {Andrews}, {Pontoppidan}, {Herczeg}, {van Kempen}  \& {Miotello}}{{van der
  Marel} et~al.}{2016}]{van-der-Marel16}
{van der Marel} N.,  {van Dishoeck} E.~F.,  {Bruderer} S.,  {Andrews} S.~M.,
  {Pontoppidan} K.~M.,  {Herczeg} G.~J.,  {van Kempen} T.,   {Miotello} A.,
  2016, \mn@doi [\aap] {10.1051/0004-6361/201526988}, \href
  {http://adsabs.harvard.edu/abs/2016A%26A...585A..58V} {585, A58}

\bibitem[\protect\citeauthoryear{{van der Marel} et~al.,}{{van der Marel}
  et~al.}{2018}]{van-der-Marel18}
{van der Marel} N.,  et~al., 2018, \mn@doi [\apj] {10.3847/1538-4357/aaaa6b},
  \href {https://ui.adsabs.harvard.edu/abs/2018ApJ...854..177V} {854, 177}

\makeatother
\end{thebibliography}




\appendix

\section{Resolution Study}

\begin{figure}
	\includegraphics[width=0.5\textwidth]{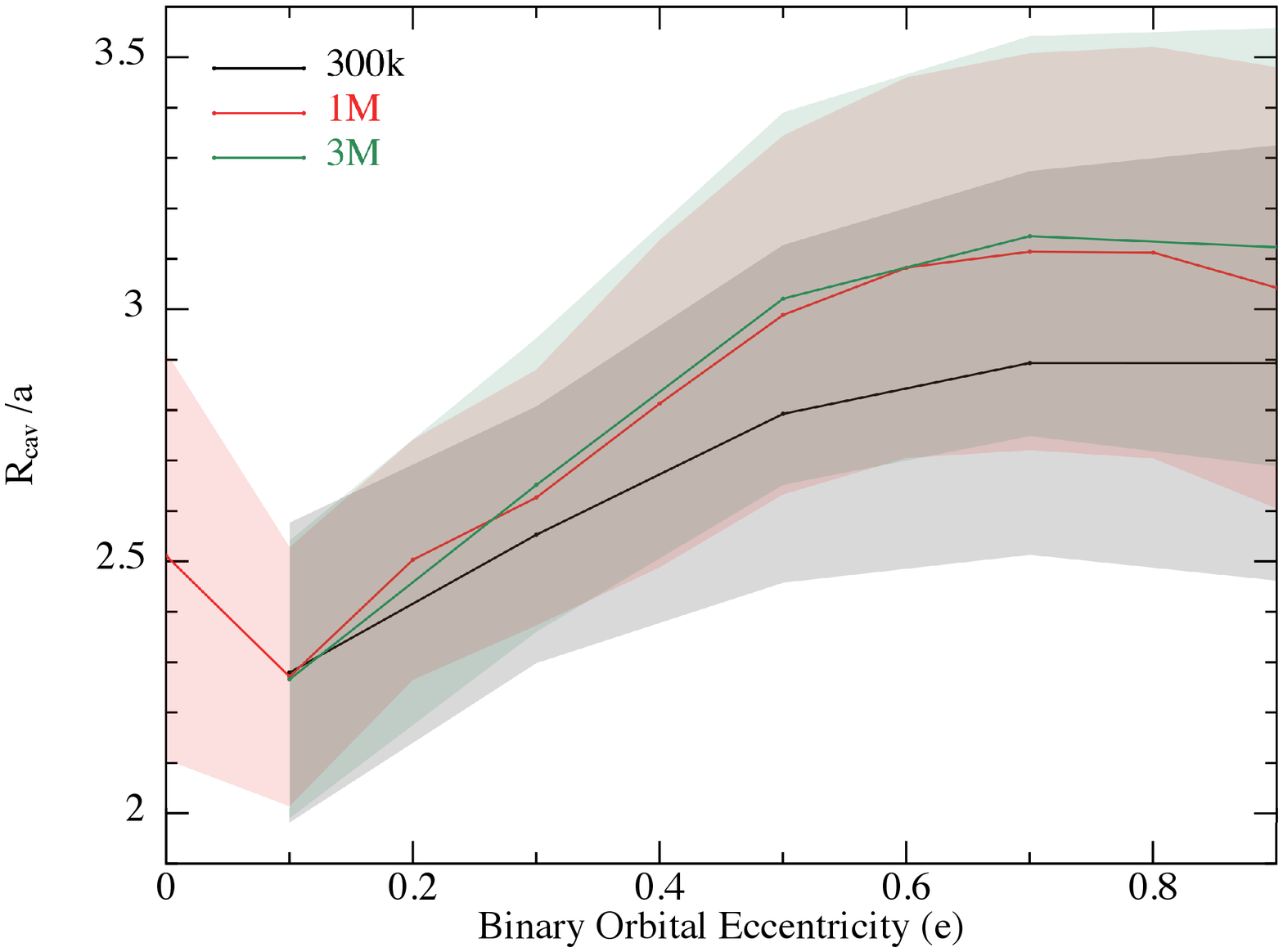}
    \caption{Cavity size as a function of binary orbital eccentricity for a coplanar disc with $(H/R)_{\rm in} = 0.05$ and $q = 0.01$. Different lines represent different number of SPH particles.
    }
    \label{fig:resolution_study}
\end{figure}

We performed a resolution study by simulating the fiducial disc (coplanar, $q = 0.1, (H/R)_{\rm in} = 0.05$) with 300 thousand and 3 million particles, and comparing to the 1 million particle case used above.
For the resolution study we only considered binary eccentricities from $e = 0.1$ to $e = 0.9$ with a step size of $0.2$.
Figure~\ref{fig:resolution_study} shows the cavity size as a function of binary eccentricity for these simulations.
The results only converge for simulations with at least 1 million particles, while with only 300 thousand particles the cavity size is underestimated.
This underestimated cavity size is due to the scale height not being resolved, leading to increased numerical viscosity.
At one million particles the scale height is resolved, so any further increase in resolution gives no improvement in the cavity size estimate.

\bsp	
\label{lastpage}

\end{document}